\documentclass[a4paper, amsfonts, amssymb, amsmath, reprint, showkeys, nofootinbib, twoside, , superscriptaddress]{revtex4-1}
\usepackage[english]{babel}
\usepackage[utf8]{inputenc}
\usepackage[colorinlistoftodos, color=green!40, prependcaption]{todonotes}
\usepackage[pdftex, pdftitle={Article}, pdfauthor={Author}]{hyperref} 
\usepackage{color}
\usepackage{subfigure}
\def\bel#1{\begin{equation} \label{#1}}

\def\vp{\varphi}
\def\mpl{M_{\rm pl}}

\def\be{\begin{equation}}
\def\ee{\end{equation}}
\def\bea{\begin{eqnarray}}
\def\eea{\end{eqnarray}}

\def\ltap{\ \raise.3ex\hbox{$<$\kern-.75em\lower1ex\hbox{$\sim$}}\ }
\def\gtap{\ \raise.3ex\hbox{$>$\kern-.75em\lower1ex\hbox{$\sim$}}\ }
\def\gl{\ \raise.5ex\hbox{$>$}\kern-.8em\lower.5ex\hbox{$<$}\ }
\def\roughly#1{\raise.3ex\hbox{$#1$\kern-.75em\lower1ex\hbox{$\sim$}}}

\def\pref#1{(\ref{#1})}

\def\mpl{M_{\rm pl}}
\def\vp{\vec{p}}
\def\varp{\varphi}

\def\mv{m_{\varphi}}

\newcommand{\comments}[1]{}
\newcommand{\nef}{\Delta N_{\rm eff}}
\newcommand{\sh}{{\hat{s}}}
\newcommand{\ben}{\begin{enumerate}}
\newcommand{\een}{\end{enumerate}}
\newcommand{\bi}{\begin{itemize}}
\newcommand{\ei}{\end{itemize}}
\newcommand{\ba}{\begin{align}} 
\newcommand{\ea}{\end{align}}

\def\beq{\begin{equation}}
\def\eeq{\end{equation}}
\def\bea{\begin{eqnarray}}
\def\eea{\end{eqnarray}}

\begin{document}
\title{Non-thermal Hot Dark Matter from Inflaton/Moduli Decay: The Momentum Distribution and  Relaxing  the Cosmological Mass Bound}

%
%

\author{Sukannya Bhattacharya}
    \email[]{sukannya@prl.res.in}
    \affiliation{Theoretical Physics Division, Physical Research Laboratory, Navrangpura, Ahmedabad - 380009, India.}
\author{Subinoy Das}
    \email[]{subinoy@iiap.res.in}
    \affiliation{Indian Institute of Astrophysics, Bengaluru, Karnataka, 560034, India.}
\author{Koushik Dutta}
    \email[]{koushik.physics@gmail.com}
    \affiliation{Indian Institute of Science Education And Research Kolkata, Mohanpur, WB 741 246, India.}
    \affiliation{Theory Division, Saha Institute of Nuclear Physics, HBNI, Kolkata- 700064, India.}
    \author{Mayukh Raj Gangopadhyay}
    \email[]{mayukh@ctp-jamia.res.in}
    \affiliation{Centre for Theoretical Physics, Jamia Millia Islamia, New Delhi 110025, India.}
    \author{Ratul Mahanta}
    \email[]{ratulmahanta@hri.res.in}
    \affiliation{Harish-Chandra Research Institute, HBNI, Allahabad 211019, India.}
    \author{Anshuman Maharana}
    \email[]{anshumanmaharana@hri.res.in}
    \affiliation{Harish-Chandra Research Institute, HBNI, Allahabad 211019, India.}


\begin{abstract}
Decay of the inflaton or moduli which dominated the energy density of the universe at early times leads to
    a matter to radiation transition epoch. We consider non-thermal
sterile dark matter particles produced as decay product during such transitions.  The particles have a characteristic energy distribution - that associated with decays
taking place in a matter dominated universe evolving to radiation domination. We primarily focus on the case when the particles are hot dark matter, and study  their effects on the
 Cosmic Microwave Background (CMB) and Large Scale Structure (LSS), explicitly taking into account their non-thermal momentum distribution. Our results for CMB angular power  and  linear matter power spectra reveal interesting features - such as an order of magnitude higher values of hot dark matter mass in comparison to the thermal case being consistent with the present data. We observe that this 
is related to the fact that $\Delta N_{\rm eff}$ 
and the hot DM energy density can be independent of each other unlike the  case of thermal or non-resonantly produced sterile hot DM.  We also find  features in the CMB at low $\ell$ angular power potentially related to
 supersonic transmission of hot dark matter through the photon-baryon plasma. 
\end{abstract}


\maketitle

\section{Introduction}
\setcounter{footnote}{0}
\label{sec:intro}

 Understanding the nature of dark matter is a central question in both particle physics and cosmology. The physics 
 of a constituent species of dark matter
not only depends on its mass and interactions but also on its momentum distribution function. For species that
thermalize, the momentum distribution is either Fermi-Dirac or Bose-Einstein. On the other hand, for non-thermal
constituents the momentum distribution crucially depends on their production mechanism.
Thus, as we enter the era of precision cosmology, it is important to isolate natural production mechanisms for dark matter constituents,
the associated  momentum distributions  and explore their implications \cite{Bertone:2010zza}. 

From the point of view of theoretical models, it is natural for the early universe to enter
an epoch of matter domination. The inflationary paradigm has emerged as the leading candidate
for providing an explanation for the fluctuations in the CMB and the matter power spectrum. In this
context, if the inflaton decays perturbatively, then the reheating epoch is matter dominated with 
cold inflaton particles dominating the energy density of the universe \cite{Mukhanov:2005sc}. Furthermore,
in string and supergravity models, an epoch of early matter domination arising from vacuum misalignment
of moduli fields is a generic feature \cite{mod1,fq,banks, Cicoli:2016olq} (see e.g \cite{Kane:2015jia, amin} for reviews). 
An epoch of matter domination ends with the decay of the associated cold particles.
This decay process effectively provides a set of initial conditions for the evolution of the 
universe. For decay products that thermalize, thermalization leads to loss of all information about the 
kinematics of the decay process. But in a setting  with a large number of hidden sectors
(which is the generic expectation in string theory \cite{jim}) one can expect that some
of the species produced during the decay do not thermalize due to very weak interactions. In this case, the energy distribution
of the species takes a characteristic (non-thermal) form -- that associated with the kinematics of decays
taking place in a matter dominated universe evolving into a radiation dominated universe (with
the matter to radiation transition taking place as a result of the decay). This energy distribution
from decays in such transition epochs has been studied in the context of primordial nucleosynthesis in \cite{oturner}, in the context of moduli decaying to light axions in \cite{cm}.  We consider the scenario where the decaying particle decays to the Standard Model sector and the sterile particles. But the sterile particle is decoupled  from the SM plasma from the very beginning and keeps free-streaming all the way to the present epoch. 

As described above, a well-motivated setting for the production of non-thermal constituents is during the
matter to radiation transition epoch, with sterile particles being one of the decay products.  The goal of this paper is to study the 
precise  implications for the Cosmic Microwave Background (CMB) and Large Scale Structure (LSS) of 
sterile hot dark matter produced by this mechanism.
A key input for this is the momentum
distribution today of the sterile particles produced. We  obtain this from the energy
distribution computed in \cite{oturner, cm}. The momentum distribution depends on the
mass of the decaying particle, its half life and the branching ratio for the decay to the sterile particles. 
In this paper, we treat these quantities and the mass of the sterile particles produced as phenomenological parameters 
while studying the cosmological implications making use of the publicly available package CLASS \cite{class101, class102, cl4}.

We will focus on the case in which the sterile particles produced from the decay of inflaton/moduli act as hot dark matter that constitutes a small fraction of the total dark matter density today. Recall that though a sterile particle/neutrino with higher mass $m \gtrsim 5\, {\rm keV} $ is a viable  warm  dark matter candidate \cite{Dodelson:1993je,Valdarnini:1998zy,Boyarsky:2008mt,Bezrukov:2009th,Petraki:2007gq,Laine:2008pg,Abazajian:2005xn,Irsic:2017ixq,Abazajian:2019ejt,Samanta:2020gdw}, but a lighter fermion (hot dark matter) with $m \sim {\rm eV}$ usually falls into the  dark-matter misfortune as eV mass particle  free-streams until relatively late times in cosmic evolution and erases structure on small scales.  Only a small fraction of the dark matter abundance can be in the form of neutrinos or other light species with $m \sim {\rm eV}$ and this puts a stringent upper bound on neutrino mass
\cite{pl18} \footnote{But as the neutrino has all the relevant properties of DM, except free-streaming, there have been efforts to revive the neutrino or lighter sterile neutrino as viable dark matter candidate with nontrivial cosmological histories or exotic interactions \cite{Das:2020wfe,Das:2006ht}.}.
On the other hand, recent data from MiniBooNE experiment
might indicate the existence of light sterile neutrino states of $\sim (1 - 10)$ eV 
\cite{Aguilar-Arevalo:2018gpe}. Within the “3+1” neutrino oscillation framework, these results are, however, very difficult to reconcile with the absence of anomalies in the $\nu_{\mu} \rightarrow \nu_{\mu}$ disappearance as probed by recent atmospheric\cite{Aartsen:2017bap}
and short baseline \cite{Adamson:2017zcg}
experiments. If these results are confirmed by future analyses, it is likely that new physics beyond the (sterile + active) oscillation models would be necessary to resolve the tension between neutrino appearance and disappearance data.

Cosmology provides a complementary means to probe \rm{eV} scale neutrino/hot DM particles. Cosmological observables such as the CMB and large-scale structure (LSS) are also sensitive to the presence of new interactions \cite{Kreisch:2019yzn, Dasgupta:2013zpn} in the neutrino sector that would modify their standard free-streaming behavior during the radiation-dominated epoch.
As shown in \cite{Kreisch:2019yzn}, this would change the hot DM mass bound as well put constraints on effective radiation degree of freedom $\Delta N_{eff}$.  Another  important factor is whether the light sterile neutrinos are  fully thermalised or not. Not only that a non-thermal distribution function would have implications for short baseline anomaly \cite{Dentler:2019dhz} but also hot DM mass bound and its effective contribution  to the radiation energy density would change considerably \cite{Dodelson:2005tp,s2, Hannestad:2012ky, Oldengott:2019lke}.  

 For our analysis, we consider  what seems to us as a simple and well-motivated setting. The
decaying particle $(\varp)$ is the inflaton with mass $m_{\varp}~\sim~10^{-6}M_{\rm pl}$. We take its lifetime
to be of the order of $10^{8}/m_{\varp}$ to $10^{9}/m_{\varp}$. This can arise if the decay takes place via
a non-renormalisable interaction at  approximately the GUT scale.   Interestingly, we find that in this regime of parameter space the 
mass of this candidate hot dark matter particles can be significantly higher than the standard thermal hot dark matter case and still be consistent
with data. Another interesting aspect is the $\ell$ dependence of the effects on the CMB. For large $\ell$, the main effect is from $\Delta N_{\rm eff}$ as it changes the Hubble expansion rate prior to photon decoupling  which changes the silk damping scale \cite{1104.2333}. As our hot DM particle increases $\Delta N_{\rm eff}$, we also see the expected suppression in power on large  $\ell$ CMB angular power spectra. It is instructive to note that we fix our decay parameters to the range of values which obey Planck bound of $\Delta N_{\rm eff}$.
For lower values of $\ell$, the effects
due to supersonic transmission of hot dark matter through the photon-baryon plasma can be important \cite{Kreisch:2019yzn}.   Future MCMC analysis (work in progress) will make these effects more clear and would shed light whether one can detect these subtle effects of non-thermal hot DM produced from early decay through future CMB experiments. 

The subject of hot dark matter in cosmology has a vast body of literature, the reader might find the 
papers \citep{jj, j2, Choudhury:2018sbz, astro-ph/0404585,hep-ph/0402049,Oldengott:2019lke,astro-ph/0502465,Dodelson:2005tp,astro-ph/0607086,hep-ph/0012317,astro-ph/0403323,hep-ph/9303287,astro-ph/9505029, 0803.2735,hep-ph/0504059,0803.1585,astro-ph/0302337,hep-ph/0412181,astro-ph/0105385,astro-ph/0412066,astro-ph/0503612,astro-ph/0309135,astro-ph/0403291,astro-ph/9903475,1808.05955,1712.01857,1509.07471,1803.07561,hep-ph/0312154,1507.02623,1406.2961,1310.1774,1303.6267,1303.5379,1302.2516,1303.0143,astro-ph/0507544,1212.1472,1111.0605,1111.6599,1309.5383,0808.3137,0705.2406,1511.00975,Hamann:2013iba,Hamann:2011ge,Hannestad:2005bt} and the references therein interesting in the context of the present work. More specifically,
the paper \cite{jj} initiated the study of hot/warm dark matter from decays. For general overviews, see  e.g \cite{Lesgourgues:2014zoa, hep-ph/0202122, 1302.1102,1407.0017, 1309.5383,Abazajian:2017tcc}. 
Before closing the introduction, we would like to reemphasise the novel aspects of the present work. We 
study the implications on the CMB and LSS, explicitly incorporating the form of the momentum distribution
of sterile particles produced as a result of decays taking place in the matter to radiation transition (it is natural
for the universe to enter such an epoch during perturbative decay of the inflaton or as a result of vacuum misalignment
of moduli). The numerical analysis is carried out using CLASS, and it reveals some interesting features: for the hot dark matter
produced from decay of the inflaton, considerably higher (compared to the thermal case) hot DM mass can be allowed by CMB and LSS observations. Furthermore, deviations at
 low $\ell$ which can be associated with supersonic transmission of hot dark matter through the photon-baryon plasma
 are seen.

This paper is structured as follows. In section \ref{distribute}, we obtain the momentum distribution function of sterile
 particles produced from decay during a matter to radiation transition epoch. In section \ref{coseff}, we begin by briefly reviewing
 the effects that hot dark matter can have on the CMB and LSS. We then go on to input the momentum distribution function
 to CLASS and obtain our results. Finally in section \ref{disc}, we discuss future directions and conclude.
\section{Sterile decay products and the associated momentum distribution function}
\label{distribute}
\comments{

 \subsection{The Instantaneous Decay Approximation}

    The cosmology associated with a heavy scalar which dominated the energy density of the universe is often studied
in the instantaneous decay approximation. In this approximation, one takes the universe to be dominated by cold
$\varp$ particles (which behave as matter) until $t = \tau = {1 \over \Gamma_{\varp} }$, at $t = \tau$ all the $\varp$ particles decay (where $\Gamma_{\varp}$ is the width of $\varp$). A consequence of this approximation is that all the sterile particles have the same momentum throughout the cosmological history. This is because they are all produced at the same time $(t = \tau)$
and they all have the same energy $(= \mv/2)$ at the time of their production. Thus the momentum distribution function is a delta function.This makes the instantaneous approximation
unsuitable for studying the effect of the sterile particles on the CMB and LSS. Yet, we begin by discussing 
some basic features of the cosmology in the instantaneous decay approximation (see e.g  \cite{Kolb, fq, dr1, cm} for more details), as this will  give us an idea about some of the important scales involved in the problem.

     In a matter dominated epoch, the Hubble parameter: $H(t) = \left( {2 \over 3t} \right) $. Thus the energy density
at the time of decay is
  \bel{rid}
     \rho^{\rm id} = 3 \mpl^2 H^2(\tau) = { 4\mpl^2 \over 3\tau^2},
  \ee
where the superscript {\rm id}  indicates that this is a quantity computed in the instantaneous decay
approximation. This implies that the number density of $\varp$ particles at the time of decay is
\bel{nid}
  n^{\rm id}_{\varp}  =  { \rho^{\rm id} \over \mv} =  { 4\mpl^2 \over 3\tau^2  \mv}.
 \ee
The number density of the sterile particles right after the time of decay is
\bel{drid}
  n_{\rm sp}^{\rm id} = 2 B_{\rm sp} n^{\rm id}_{\varp} = { 8 B_{\rm sp} \mpl^2 \over 3\tau^2  \mv},
 \ee
where $B_{\rm sp}$ is the branching fraction to the sterile particles (the factor of 2 arises as when a $\varp$ particle decays to a sterile particle, it decays to two of them). The Standard Model sector thermalises, its reheating temperature is
\bel{rhid}
  T^{\rm id}_{\rm rh} = \left( { {40(1 - B_{\rm sp}) }\over{ \pi^2 g_{*}(T^{\rm id}_{\rm rh})} }  \right)^{1/4} \left(M_{\rm pl} \over \tau\right)^{1/2}.
\ee
 In order to compute the abundance $(Y)$ of sterile particles, we need the entropy density of the Standard Model sector at the time of decay (the number density has been already obtained in \pref{drid}). This is given by
 \bel{sid}
     s^{\rm id} ={4 \over 3} . { 2^{5/4} \pi^{1/2} \over 3. 5^{1/4} } g_{*}^{1/4} (T_{\rm rh}^{\rm id}) (1 - B_{\rm sp})^{3/4}  \left(  { \mpl \over \tau} \right)^{3/2}.
   \ee
Combining this with \pref{drid}, we find
   \bel{yid}
    { Y^{\rm id} } = { n^{\rm id}_{\rm sp} \over s^{\rm id} } = {3 \over   \pi^{1/2}  }. \left( {5 \over 2} \right)^{1/4} { B_{\rm sp} \over {(1 - B_{\rm sp})^{3/4}}} {1 \over  g_{*}^{1/4}(T^{\rm id}_{\rm rh})}
    \left( { \mpl \over \mv^2 \tau} \right)^{1/2}.  
    \ee

As mentioned earlier, although we will not be using the instantaneous decay approximation to compute the
effect of the sterile particles on the CMB and the LSS, the above quantities will set the scale for various physical
quantities in our computations.
}
     
      In our scenario, the sterile particles are produced from the decay of a heavy massive scalar $(\varphi)$
We consider $1 \to 2$ decays, with identical decay products. The
 production takes place in the early universe, when $\varp$ is decaying and the universe is in a matter to radiation transition
 epoch. The species
 $\varp$ decays to the sterile particle with branching ratio $B_{\rm sp}$, these particles do not thermalise. The remaining
 decay products thermalise (as this sector contains the Standard model, we will refer to this as the Standard Model sector).    
     

   The momentum  distribution of the sterile particles is central to obtain their effect on the CMB and LSS.
To compute this momentum distribution, one needs to know the scale factor of the universe during the
epoch that $\varp$ decays. Thus, we start by discussing the evolution of the scale factor during this epoch
\cite{cm, Mukhanov:2005sc}.
 
 \subsection{The Scale Factor}
 \label{Ssf}
 
\begin{figure}[h]
\centering
  \begin{minipage}{.46\textwidth}
  \centering
    \includegraphics[width=\textwidth]{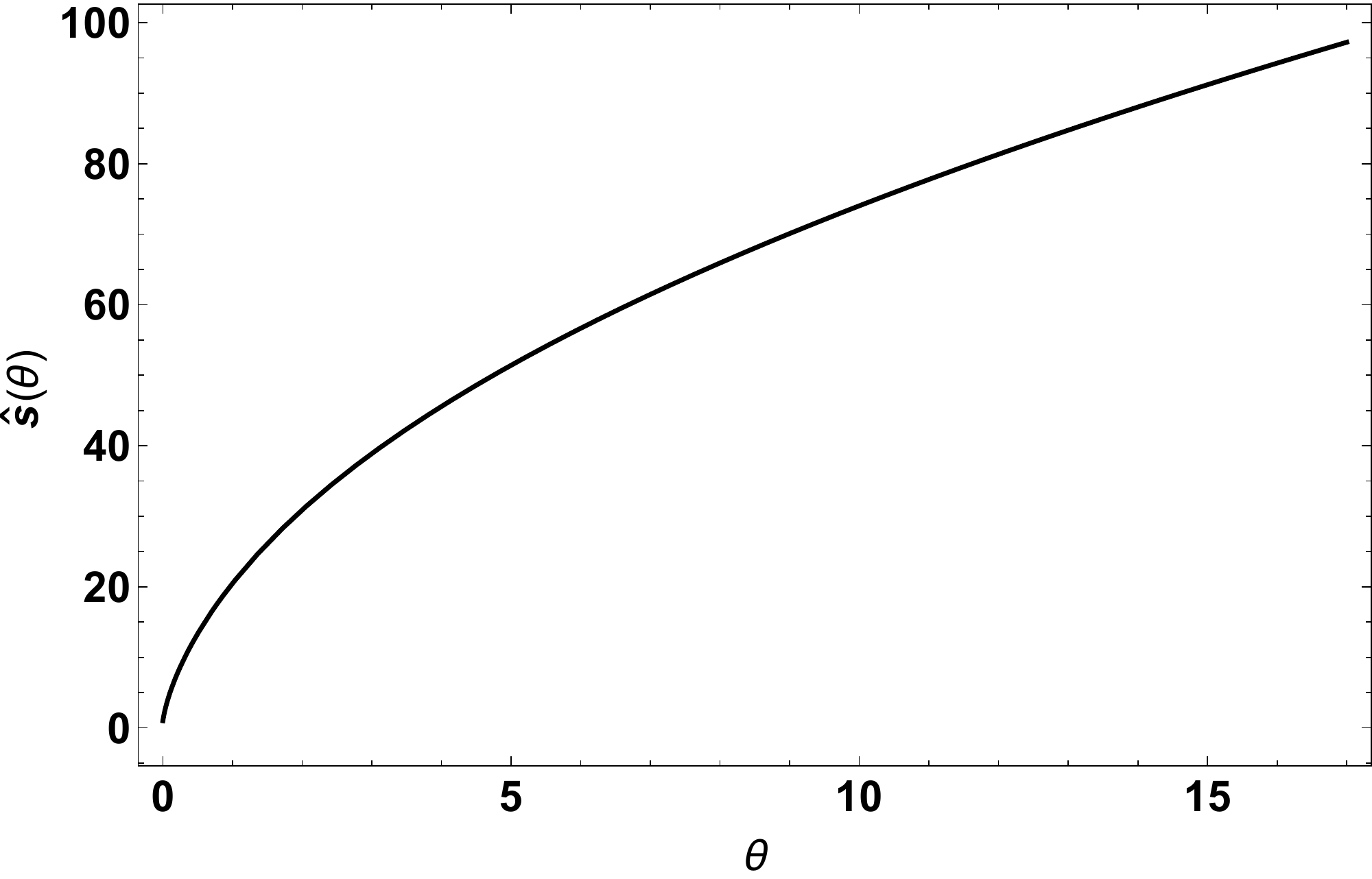}
    \caption{ The scale factor as a function of the dimensionless time.}
    \label{figS}
  \end{minipage}\qquad
  \begin{minipage}{.46\textwidth}
  \centering
    \includegraphics[width=\textwidth]{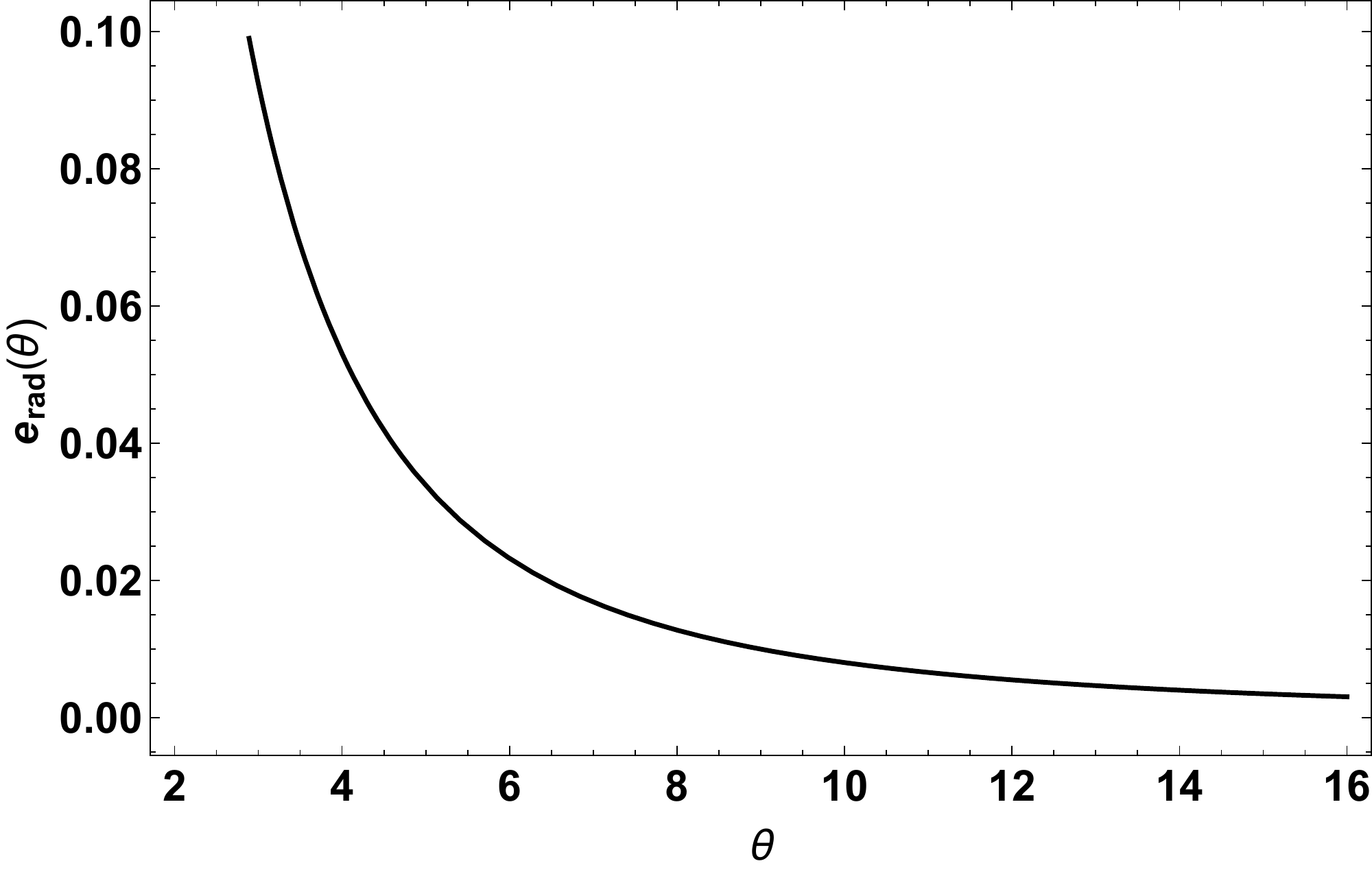}
    \caption{The energy density in radiation  as a function of the dimensionless time}
    \label{frad}
  \end{minipage}
\end{figure}

   The evolution of the universe during the epoch that $\varp$ decays is governed by the equations:
\bel{matcon}
  \dot{\rho}_{\rm mat} + 3 H \rho_{\rm mat} = - { \rho_{\rm mat} \over \tau},
\ee
\bel{radcon}
  \dot{\rho}_{\rm rad} + 4H \rho_{\rm rad} = + {\rho_{\rm mat} \over \tau},
\ee
and
\bel{evo}
    H = \left( { \dot{a} \over a } \right) = \sqrt{ { { \rho_{\rm mat} + \rho_{\rm rad} } \over 3 M^2_{\rm pl} }}.
\ee    
 In the above,  $\rho_{\rm mat}$ denotes the energy density in matter and $\rho_{\rm rad}$ is the energy density
 in radiation. The energy density in radiation is  the sum of the energy densities in the Standard Model sector and sterile particles (since the
 sterile particles are highly relativistic at the time of production, thus they contribute to the energy density as radiation
 during the epoch that $\varp$ decays). It is useful to introduce the dimensionless variables
 \begin{align}
 &\theta = { t \over \tau}, \ \ \  &&\sh(\theta) = a (\tau \theta), \ \ \ \textrm{and}  \nonumber \\
  e_{\rm mat}(\theta) = &{ \tau^2 \rho_{\rm mat} (\tau \theta) \over M_{\rm pl}^2 }, \ \ \ &&e_{\rm rad}(\theta) = { \tau^2 \rho_{\rm rad}  (\tau \theta) \over M_{\rm pl}^2 }.
  \label{sdef}
 \end{align}
 %
 Now, let us come to the initial conditions. We will take the starting point of our numerical evolutions to be $t=0,$ and
 work with conventions in which the scale factor is equal to unity at this point. At this stage, the universe is completely matter dominated, thus we will take
\comments{
\bel{matinn}
   \rho_{\rm mat}(0) = \alpha \rho^{\rm id} \phantom{abc} \textrm{i.e}   \phantom{abc} e_{\rm mat}(0) = {4 \over 3} \alpha
 \ee
}
\bel{matinn}
   \rho_{\rm mat}(0) = {4 \alpha \over 3}  {M^2_{\rm pl} \over \tau^2 } \phantom{abc} \textrm{i.e}   \phantom{abc} e_{\rm mat}(0) = {4 \over 3} \alpha,
 \ee
with $\alpha \gg 1$ (the factor of 4/3 has been chosen so as to get some numerical simplifications) and the energy density in radiation to be zero. We note that the solution will be universal\footnote{This is a consequence of the fact that for energy densities much greater than ${ \mpl^2 \over \tau^2}$, the Hubble time is much smaller than
$\tau$.}  in the sense that we will get the same late
time universe as long as $\alpha \gg 1$. In our numerics, we will take $\alpha = 10^{4}$.  We exhibit the results of numerical integration the evolution equations in the form of plots. 
 Figure \ref{figS}, shows the
evolution of the scale factor while figure \ref{frad} exhibits  the energy density in radiation as a function of the dimensionless time $\theta$.

\subsection{The energy distribution at early times}

  The time scale for the decay of the $\varp$ particles is $\tau$, at these early times the sterile particles produced
are highly relativistic. Let us begin by discussing their energy distribution at these early times as obtained in \cite{oturner,cm}. 
The treatment in \cite{oturner} is rather brief (the final result of the computation
is presented as a plot), we will primarily follow \cite{cm}.
The comoving modulus number
density  as function of time is given by
\bel{mnd}
  N = N(0) e^{-t/ \tau},
\ee
where $N(0)$ is the number density at $t=0$. Note that $N(0)$ can be computed from \pref{matinn}.
\bel{nnin}
  N(0) = {\rho_{\rm mat}(0) \over \mv} = { 4 \alpha \mpl^2 \over {3\tau^2  \mv}}.
 \ee
If a sterile particle is produced from decay of a modulus at time $t=t_d$, then at that point of time
it has energy  $\hat{E} = {\mv/2}$. At a later time t, its energy is given by\footnote{In this subsection
we will assume that t is small enough so that  the sterile particles continue to be highly relativistic
at t.}
$$
  E = \hat{E}  \left( {a(t_d)  \over a(t)} \right).
$$
Thus at time $t$, sterile particles produced between $t_d$ and $t_d + dt_d$ have energies
in the range
\bel{eran}
  dE = E H(t_d) dt_d.
 \ee
The number density of sterile particles produced between  $t_d$ and $t_d + dt_d$ can be computed from \pref{mnd}:
\bel{ns}
 dN = {2 B_{\rm sp} \over \tau} N(0) e^{-t_d/\tau} dt_{d}.
\ee
Equation \pref{eran} and \pref{ns} can be combined to compute the co-moving number density spectrum
(as a function of the energy) at time $t$ \cite{cm}:
\bel{com}
 dN_t^c = {2 B_{\rm sp} \over \tau} N(0) e^{-t_d/\tau} {1 \over E H(t_d)} dE.
 \ee
where  $t_d$ is to be expressed in terms of t and $E$, by making use of the relation $  E= \hat{E} { a(t_d)  \over a(t) }$.
The physical number density is obtained by dividing this by $a^{3}(t)$. Doing this, and converting to our dimensionless
variables introduced in \pref{sdef} one obtains the spectrum at time $t$ to be:
\bel{Nap}
 dN_{t}  =
  {1 \over \sh^{3}(\theta) }2 N(0) B_{\rm sp} {e^{-\sh^{-1}(y) }\over { \hat{H} (\sh^{-1} (y) ) E}} dE \equiv \tilde{n}_{t}(E) dE.
\ee 
where we have introduced the variable $y$:
 \bel{yd} 
  y \equiv { E \sh (\theta) \over \hat{E}},
 \ee 
 with $\theta = t/ \tau$. $\hat{H}$ is the dimensionless Hubble parameter  $\hat{H} \equiv {\sh'(\theta) / \sh(\theta)} $.
 Finally, $\sh^{-1}$ is the inverse of the scale factor function introduced in \pref{sdef}. We note that the spectrum in \pref{Nap} is non-vanishing for E in the range ${ \hat{E}  \over \sh(\theta)}  < E <  \hat{E}$. The lower limit corresponds to decays at the initial time and upper limit corresponds to decays that at occur at $\theta$.
 This implies that $y$ varies between 1 and $\sh(\theta)$.

\subsection{The momentum distribution function today}

  The discussion in the previous subsection can be used to compute the momentum distribution of the
sterile particles today. At early times, the sterile particles are highly relativistic. The momentum
distribution can be computed from the energy distribution in \pref{Nap} using isotropy.
\bel{mnum}
  dN_t =  {  \tilde{n}_t (|\vec{p}|) \over 4 \pi |\vec{p}|^2 } d^3p \equiv n_t(\vec{p}) d^{3} p. 
\ee
The momentum distribution  today can be obtained by making use of the fact that after production the sterile
particles free stream. Thus if $t^{*}$ is an early  time such that almost all the $\varp$ particles have decayed by $t^{*}$,
then the momentum distribution of the sterile particles  today is given by
\bel{ndter1}
   n_{t_0}( \vp) =   n_{t^*} \left( { a(t_0)  \over a(t^{*}) } \vp\right). 
 \ee
 Note that given our earlier discussion regarding the values of the argument of $\tilde{n}_{t}$ for which it is non-vanishing,  $n_{t_0}( \vp)$ is non-vanishing if
 \bel{prange}
  { \hat{E} \over a(t_0) } <   |\vec{p}|  <   { {\hat{E} a(t^{*}) }\over a(t_0) }.
 \ee 
 For our numerics, we shall use $t^{*} = 15 \tau$ (i.e $\theta^* =15$).

    The CLASS routine requires that the momentum distribution is expressed in units of   $T_{\rm ncdm,0}$,    
 the typical momentum of the dark radiation particles today.     Equation \pref{prange} gives the range of the momentum for the sterile particles. The range of momentum in \pref{prange} corresponds to
 decays taking between $t = 0$ and $t=15 \tau$, we expect most of the decays to take place in the early part
 of this range in time.  This motivates us to take
\bel{tmom}
  T_{\rm ncdm,0} = { \hat{E} \over 4 }{ a(t^*) \over a(t_0) } = {1 \over 8} \mv \left( { g(t_0) \over g(t^*) } \right)^{{1\over 3}} \left(  { T(t_0) \over T(t^*) } \right).
\ee
$T(t^*)$, the temperature of the Standard Model sector  at $t^*$ can be computed from numerical 
analysis of the evolution of the energy density in radiation carried out in section \ref{Ssf}. The energy density in the Standard Model sector at $t^*$ is
\comments{
\bel{rts}
\rho_{\rm sm} (t^*) =  { \mpl^{2} \over \tau^{2} }(1 -  B_{\rm sp}) e_{\rm rad} (\theta^{*}) = {3 \over 4} \rho^{\rm id}_{\rm sm} e_{\rm rad} (\theta^*), 
\ee
}
\bel{rts}
\rho_{\rm sm} (t^*) =  { \mpl^{2} \over \tau^{2} }(1 -  B_{\rm sp}) e_{\rm rad} (\theta^{*}), 
\ee
Thus the temperature of the plasma at $t=t^{*}$ is

\bel{tts}
   T(t^{*}) =   \left( {3 \over 4} e_{\rm rad} (\theta^{*}) \right)^{1/ 4}  \left( { {40(1 - B_{\rm sp}) }\over{ \pi^2 g_{*}(T(t^*))} }  \right)^{1/4} \left(M_{\rm pl} \over \tau\right)^{1/2}.
\ee
\comments{
where $T_{\rm rh}^{\rm id}$ is as given in \pref{rhid}. . Thus $T_{\rm rh}^{\rm id}$ and $T(t^{*})$ differ approximately by a factor of five,
we will
take that there is no change (or a change of negligible effect) in $g_{*}$  between these temperatures.}
 Our numerics in section \pref{Ssf} give $\left( {3 \over 4} \right)^{1 / 4}  ( e_{\rm rad}( 15  ) )^{1/4} = {{0.2262}}$. Now, taking $g(t_0) = 3.91$ and $g(t^*) \sim 100$ in \pref{tts}, we obtain
\bel{tmomf}
  T_{\rm ncdm,0}  =  0.418 \left( \mv^2 \tau \over M_{\rm pl} \right)^{1/2} { T_{\rm cmb}  \over {(1 - B_{\rm sp})^{1/4} }} \equiv \zeta T_{\rm cmb},
\ee
where we have defined $\zeta =  { 0.418 \over  {(1 - B_{\rm sp})^{1/4} }} \left( \mv^2 \tau \over M_{\rm pl} \right)^{1/2} $.
Finally, we express the momentum distribution function in terms of the momentum is units of $T_{\rm ncdm,0}$,
 \bel{dless}
   q \equiv {|\vec{p}| \over T_{\rm ncdm,0}}.
 \ee
 Making this variable change, \pref{mnum} gives the momentum distribution in units of $(T_{\rm ncdm,0})^3$ to be 
 \bel{fq}
   f(q) =  {  32 \over { \pi \hat{E}^3 } } \left( { N(0) B_{\rm sp}  \over { \sh^{3}(\theta^{*} ) } } \right)
     { e^{- \sh^{-1}(y)} \over q^{3} \hat{H}(\sh^{-1}(y))},
 \ee
  where 
  \bel{yyyfd}
     y = {q \over 4} \sh(\theta^*),
  \ee
  and the range of q is given by
\bel{qran}
   {4 \over \sh(\theta^{*}) } < q < 4.
  \ee

 \section{ Effects on Cosmological Observables}
 \label{coseff}
  In this section, we carry out our analysis on the cosmological observables making use of the above discussed non-thermal
momentum distribution function. Let us begin by briefly reviewing the key effects that such particles can have on 
cosmology. For a more detailed discussion see e.g \cite{Lesgourgues:2014zoa}.

\subsection{Review of effects of Hot DM on cosmology}

Hot DM  Neutrino like particles have significant effect on  the expansion rate during the cosmological era when the Universe is  radiation dominated . Their contribution to the total radiation content can be parametrized in terms of  $N_{eff}$. 
Other than changing the expansion rate of the Universe, another important effect is free-streaming of hot DM until they turn non-relativistic. The physical effect of free-streaming is to damp small-scale 
density fluctuations: hot DM cannot be confined into  regions smaller than their free-streaming length, because their velocity is greater than the escape velocity from gravitational potential wells on those scales.
On the other hand, on scales much larger than the free-streaming scale, their
 velocity can be effectively considered as vanishing, and after
the non-relativistic transition the hot DM perturbations behave like
CDM perturbations. 

\subsubsection*{Effect of hot eV mass dark matter  on matter power spectrum }
On large scales (i.e on wave-numbers smaller than  $k_{nr}$),   the matter power spectrum $P(k,z)$ can be shown to depend only on the matter density fraction $\Omega_m$ today (including neutrinos which was hot earlier but now behaves like CDM). If the hot DM mass is varied with $\Omega_m$ fixed, the large scale power spectrum remains invariant but
on small scales $k >k_{nr}$, the matter power spectrum is affected by hot DM masses for essentially three reasons:

\begin{enumerate}

\item Massive hot DM does not cluster on those scales. The matter power spectrum can be written as,
\begin{align}
P(k,z) &= \left\langle 
\left| \frac{\delta \rho_{\rm cdm} + \delta \rho_{\rm b} + \delta \rho_{sp}}
{\rho_{\rm cdm} + \rho_{\rm b} + \rho_{sp}}\right|^2 \right\rangle \nonumber \\
&=  \Omega_{\rm m}^{-2} \left\langle \left| 
\Omega_{\rm cdm} \, \delta_{\rm cdm} 
+ \Omega_{\rm b} \, \delta_{\rm b} 
+ \Omega_{sp} \, \delta_{sp}\right|^2
\right\rangle~.
\end{align}
where $\delta \rho_{sp}$ and $\Omega_{sp}$ represents density fluctuation and fractional energy density of our sterile particle ( hot DM).
On scales of interest and in the recent universe, baryon and CDM fluctuations are almost equal to each other, while $\delta_{sp} \ll \delta_{\rm cdm}$. The power spectrum would be reduced by a factor $(1-f_{sp})^2$ with
\begin{equation}
f_{sp} \equiv \frac{\Omega_{sp}}{\Omega_{\rm m}}.
\end{equation}

\item
 The redshift of radiation-to-matter equality $z_{\rm eq}$ or the baryon-to-CDM ratio $\omega_{\rm b}/\omega_{\rm cdm}$ can be slightly affected by sterile particle masses, with a potential impact on the small-scale matter power spectrum. This depends  on which other parameters are kept fixed when the sterile particle hotDM mass is varied. But matter power spectra  also can be affected by perturbative cosmology in presence of hot Dark matter.

\item
 The growth rate of cold dark matter perturbations is reduced through an absence of gravitational back-reaction effects from free-streaming hot DM. This growth rate is set by an equation of the form
\begin{equation}
\delta_{\rm cdm}'' + \frac{a'}{a} \delta_{\rm cdm} = -k^2 \psi~,
\end{equation}
where $\delta_{\rm cdm}$ stands for the CDM relative density perturbation in Fourier space, and $\psi$ for the metric perturbation playing the role of the Newtonian potential inside the Hubble radius.  The right-hand side represents gravitational clustering. The second term on the left-hand side represents Hubble friction, i.e. the fact that the cosmological expansion  slows down clustering. The coefficient $a'/a$ is given by the first Friedmann equation as a function of the total background energy density. In a universe such that all species present in the Friedmann equation do cluster, as it is the case in a matter-dominated universe with $\delta \rho_{\rm total} \simeq \delta \rho_{\rm cdm} + \delta \rho_b$ and $\bar{\rho}_{\rm total} = \bar{\rho}_{\rm cdm} + \bar{\rho}_{\rm b}$, the solution is simply given by $\delta_{\rm cdm} \propto a$: the so-called linear growth factor is proportional to the  scale factor. But whenever one of the species contributing to the background expansion (like our sterile particle) does not cluster efficiently, the CDM (as well as baryons) clusters at a slower rate.
This is why measuring linear matter power spectra put strong bounds on hot dark matter mass.

\end{enumerate}

\subsubsection*{Effect of hot dark matter on CMB }

The implications for the CMB can summarised by the following three effects:
\begin{enumerate}

\item Hot dark matter can affect the redshift of matter/radiation equality $z_{\rm eq}$ via its contribution to $\rho_{\rm m}$
and $\rho_r$ which depend on its mass.  If they are relativistic at $z_{\rm rec}$  then it is reasonable to assume that they contribute only to $\rho_{\rm rad}$,
though they could be mildly relativistic near $z_{\rm eq}$. This can modify the contribution from the early ISW effect and have an effect  on the CMB.

\item  Hot dark matter changes the expansion rate by changing the energy density,  this in turn changes the size of the sound horizon at recombination and/or the distance to last scattering.  Changes in the expansion rate can also affect the damping scale and this is one of the main effects how an excess radiation affects CMB\cite{1104.2333}.

\item Free-streaming
sterile particle (or hot dark matter) can travel supersonically through the photon-baryon plasma at early times, hence gravitationally
pulling photon-baryon wavefronts slightly ahead of where
they would be in the absence of hot DM. 
\comments{Free-streaming hot DM thus leads to a physical size of the photon sound horizon at last scattering $r_{*}$
that is slightly larger than it would be otherwise.}
As a result,  free-streaming hot DM imprints a net
phase shift in the CMB power spectra at larger
scales (smaller $\ell$), as well as a slight suppression  of the amplitude.  This
phase shift is considered to be a robust signature of the
presence of free-streaming radiation in the early Universe.

\end{enumerate}

The total relativistic energy density of the Universe at late time is paramterised by $N_{\rm eff}$, where $\Delta N_{\rm eff} = N_{\rm eff} - 3.046$ corresponds to additional dark relativistic degrees of freedom other than the three neutrino flavours of the SM. In our case, the massive sterile particle with its characteristic non-thermal distribution contributes to  dark radiation. In this case, the bound is conventionally characterised by  $m_{X, \rm sterile}^{\rm eff} \equiv \Omega_{X, \rm sterile} h^2 (94.1 \rm eV)$ and $N_{\rm eff}$, where $m_{X, \rm sterile}^{\rm eff}$ is related to the physical mass of the sterile particle and the relation differs for different models. The latest PLANCK + BAO bounds on these parameters are as follows: $N_{\rm eff} < 3.29,~~ m_{X, \rm sterile}^{\rm eff} < 0.65~ \rm eV$ \cite{pl18}. From this bound, it is clear that hot dark matter can only constitute a very small fraction of the total dark matter energy density. 
\comments{
\textbf{We note that Subinoy/Koushik cleans up. 
BOSS Ly $\alpha$ data alone provide better bounds than previous Lyman $\alpha$ results, but are still poorly constraining, especially for the sum of neutrino masses $m_{\nu}$, for which one obtains an upper bound of 1.1 eV (95 CL), including systematics for both data and simulations. Lyα constraints on ΛCDM parameters and neutrino masses are compatible with CMB bounds from the {\rm Planck} collaboration\cite{}. Interestingly, the combination of Lyα with CMB data reduces the uncertainties significantly, due to very different directions of degeneracy in parameter space, leading to the strongest cosmological bound to date on the total neutrino mass, $ m_{\nu} < 0.15 eV $ at 95 CL (with a best-fit in zero). Adding recent BAO results further tightens this constraint to $ m_{\nu} < 0.14 eV$ at 95 CL. 
But bounds relaxes with non-trivial neutrino interaction\cite{Kreisch:2019yzn} or non-trivial distribution function \cite{Acero:2008rh,Oldengott:2019lke }.
We see that in our case the bound relaxes oder of magnitude or two reason 
1. Non-thermal distribution function . a factor of (1/10)
2. $\Delta N_{eff}$ independent of  dark radiation energy density $w$ 
}
}
 
\subsection{Finding CMB and LSS observables for our non-thermal hot dark matter }

   Keeping the above effects in mind and   having obtained the momentum distribution of the sterile particles, in this section we will compute their effect
on LSS and the CMB. The full computation for this will be  done numerically by modifing the publicly available  CLASS code~\cite{class101,class102} to incorporate the new distribution function. It is important to keep in mind that while the full computation takes the momentum distribution
as  input and has to be done numerically, the effect on the CMB and LSS are primarily set by three parameters which can be easily calculated once the momentum distribution is known \cite{s2}. These are:

\begin{enumerate}

\item $\Delta N_{\rm eff}$ : The number of additional relativistic species at the time of neutrino decoupling (the
sterile particles are relativistic at this point). Current bounds require $\Delta N_{\rm eff} \lesssim 0.3$, \cite{pl18}.
  In our case, the sterile particles and the Standard Model sector are both entirely produced from the decay of $\varp$  particles. Thus, the relative energy densities of the two sectors at early times is  $B_{\rm sp} / (1 - B_{\rm sp})$. Given this,
 $\Delta N_{\rm eff}$ is easily computed by standard methods, see  e.g \cite{Kolb, dr1, ghosh, cm}. One finds\footnote{This
 assumes instantaneous thermalisation of the Standard Model sector \cite{cm}.}
 \bel{ddnn}
  \nef =  { 43 \over 7 } { B_{\rm sp} \over {1 - B_{\rm sp} }  } \left(  g_{*} (T(t_{\nu})) \over g_{*} (T(t^{*}) ) \right)^{1/3},
 \ee
where $t_{\nu}$ is the time at which neutrinos decouple.  
\item $\lambda_{FS}$: Till the epoch when hot dark matter particles turn non-relativistic, they can not be bound in gravitational potential wells of cold dark matter. The co-moving distance travelled by hot DM particles till the temperature of the universe drops below their mass is known as the free streaming length. As hot DM (which is cold at the present epoch) contributes to a fraction of entire dark matter budget today, due to this early free streaming behaviour, the Linear matter power spectra generally gets suppressed   at length scales smaller than $\lambda_{FS}$. Hot dark matter turns non-relativistic deep in the
matter dominated era. A quick estimate\footnote{CLASS computes the exact free streaming wavelength directly from
the momentum distribution, the estimate we give is only for the purposes of the present discussion.} of the minimum free streaming wavenumber in our case can be obtained following \cite{Lesgourgues:2014zoa}. We find
\bel{fse}
   k_{\rm fs} \approx 0.018 \left( { m_{\rm sp} T_{\rm \nu} \over { \zeta T_{\rm cmb} \phantom{a} 1 {\rm eV}} } \right)^{1/2} (\Omega_m h)^{1/2}
    {\rm Mpc}^{-1},
\ee
with $\zeta$ as defined in \pref{tmomf}.
\comments{
We note that the  co-moving free-streaming length of our sterile particles can be expressed
 in terms of the average velocity of the sterile particles today $v_{\rm sp}$ \cite{s2}. This in turn can be written in terms
 of $\Delta N_{\rm eff}$ and $w_{\rm sp}$,
 \bel{vss}
   v_{\rm sp} = {7 \over 8} { \pi^2 \over 15} { T_{\rm cmb}^4 h^2 \over \rho^0_c } {\Delta N_{\rm eff} \over w_{\rm sp}}.
 \ee
}

\item $w_{\rm sp}$: which is related to the current energy density of the sterile particle (the product of its
current number density and its mass). Following the conventions
of \cite{s2}, we will take
\bel{simw}
  w_{\rm sp} =  m_{\rm sp} n_{\rm sp} \left[ { h^2 \over  \rho_c^0} \right],
 \ee
where $m_{\rm sp}$ and $n_{\rm sp}$ are the mass and number density of the sterile particle. $\rho^0_c$ is the critical density today and $h$ is the reduced Hubble parameter. To compute $w_{\rm sp}$, we need to compute $n_{\rm sp}$, the number density of the
 sterile particles today. To do this, we begin by computing the abundance $(Y)$ of the particles. The energy density
 of the standard model sector at $t^{*}$ is
 \begin{align}
 \rho_{\rm sm}(t^*) &= (1 - B_{\rm sp}) \rho_{\rm rad} ( 15 \tau)\nonumber \\
                    & =  (1 - B_{\rm sp}) { M^{2}_{\rm pl} \over \tau^2}  e_{\rm{ rad}}( 15 )
 \label{rhonum}
 \end{align}
%
Thus the entropy density at this point is 
\begin{align}
 s(15 \tau) = &\left({ 3 \over 4} e_{\rm{ rad}} ( 15 ) \right)^{3/4} \left( {4 \over 3} \right)
   \left({\pi^2 \over 30} \right)^{1/4} \nonumber \\ & \times g_{*}^{1/4}(T(t^{*})) \left(  {M_{\rm pl} \over \tau} \right)^{3/2}.
   \label{snum}
\end{align}
%
 The number density at $t^{*}$ can be computed similarly. The number density at the initial time is given in \pref{nnin}
 By $t=t^{*}$, almost all the
 $\varp$ particles decay; their branching fraction to the sterile particles is $B_{\rm sp}$. Thus, to a very good
 approximation the number density of the sterile particles at $t=t^{*}$ is
\bel{n15t}
   n( 15 \tau) = { 2 B_{\rm sp} N (0 ) \over {a^{3} (15 \tau) }} ,
 \ee
Now computing the abundance by taking the ratio of \pref{snum} and \pref{n15t}, and using this to compute the
number density of the sterile particles in terms of number density of neutrinos today $(n_{\nu})$ today we find
\comments{
\begin{eqnarray}
\label{Yreal}
  Y &=&  \left( {4 \over 3} \right)^{3/4} { \alpha \over {e_{\rm rad}(15 )^{3/4} a^{3} (15 \tau)}} {3 \over   \pi^{1/2}  }. \left( {5 \over 2} \right)^{1/4} { B_{\rm sp} \over {(1 - B_{\rm sp})^{3/4}}} {1 \over  g_{*}^{1/4}(T^{\rm id}_{\rm rh})}
    \left( { \mpl \over \mv^2 \tau} \right)^{1/2}.  , \cr
  &\approx& 1.13 Y^{\rm id},
 \end{eqnarray}
 where we have made use of our numerics in section \ref{Ssf} and $Y_{\rm}^{\rm id}$ is as defined in \ref{yid}.
 The number density of the sterile particles today is
\bel{drt}
   n_{\rm sp} =  Y s_0 = { 43 \pi^{4} \over 45. 3. \zeta(3) } Y n_{\nu},
 \ee
where  $s_0$ is the entropy density of the visible sector universe today and  $n_{\nu}$ is the number
density of one species of neutrinos today. 
 Combining (\ref{yid}, \ref{Yreal}) and (\ref{drt})
}
\begin{align}
n_{\rm sp} = &1.13  \left[ { 43 \pi^{4} \over 45. 3. \zeta(3) } \right]  { 3 \over {\pi^{1/2}  g_{*}^{1/4} (T(t^{*})) } }  \left( {5 \over 2} \right)^{1/4}\nonumber \\ &\times { B_{\rm sp} \over {(1 - B_{\rm sp})^{3/4}}} \left( {\mpl \over \tau \mv^2} \right)^{1/2} n_{\nu}.
\label{idww}
\end{align}
 %
 %
 From this, the parameter $w_{\rm sp}$ (as defined \pref{simw}) is found to be
 \begin{equation}
 \label{wff}
   w_{\rm sp}     =  { m_{\rm sp} \over 94.05 {\rm eV} }   { 62.1 \over {  g_{*}^{1/4} (T(t^{*})) } }   { B_{\rm sp} \over {(1 - B_{\rm sp})^{3/4}}}  \left( {\mpl \over \tau \mv^2} \right)^{1/2}.
 \end{equation}
  \end{enumerate}

 A few comments are in order:
 \begin{itemize}
     \item In the computation of $w_{\rm sp}$, various intermediate expressions depend on $\alpha$ (the dimensionless energy density
     at the intial time). We have checked $w_{\rm sp}$ is independent of this choice for the initial energy density, as long 
     as $\alpha \gg 1$. This is in keeping with the expectation that for $\alpha \gg 1$,  the late time solution is universal.
     \item It is interesting to compare the expression \pref{wff} for $w_{\rm sp}$ with that for the same in the instantaneous decay approximation,
     the primary focus of \cite{jj, j2}. While
     the functional dependence on the various parameters are the same, the overall coefficient is greater by a factor of approximately
     fifteen percent. This exhibits the importance of incorporating the exact background and the associated  distribution function.
    
 \end{itemize}
 
\subsubsection*{Results of Numerics}
CLASS (Cosmic Linear Anisotropy Solving System)~\cite{class101} is a numerical code which simulates the evolution of the background and perturbations of the universe working to the linear order. The inputs for the default code are the present day values of different cosmological parameters ($\Omega_b^{(0)}h^2$, $\Omega_c^{(0)}h^2$, $H_0$, primordial parameters from inflation etc.) for the 6-parameter $\Lambda$CDM model and its extensions. The outputs are typically the observables for CMB and LSS experiments, i.e., the temperature power spectrum of the  CMB ($C_{\ell}^{TT}$), temperature-polarization power spectra ($C_{\ell}^{TE}$) and the matter power spectrum ($P(k)$) etc. 
\begin{figure}[h]
\centering
\includegraphics[width=0.49\textwidth]{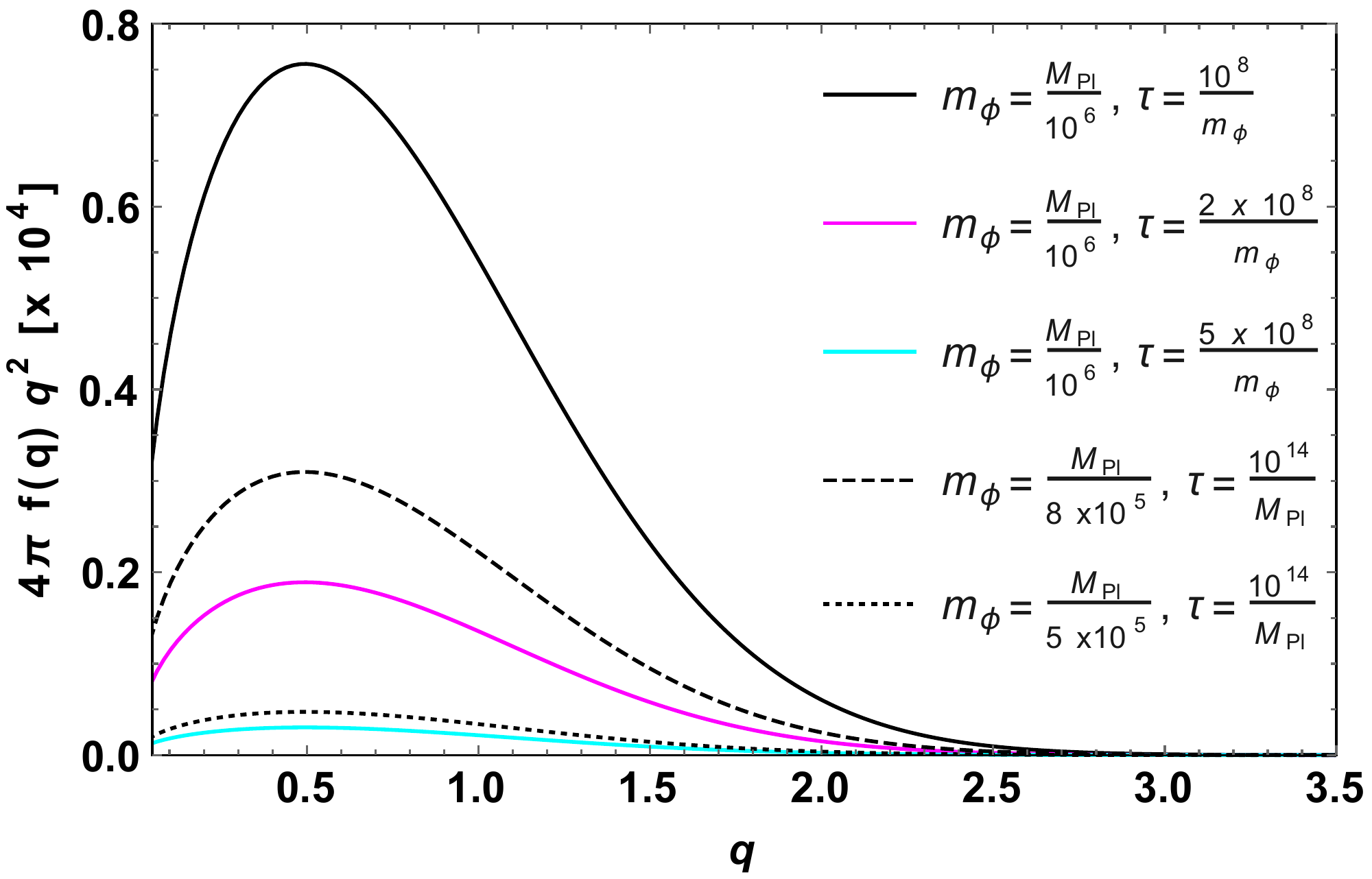}
  \caption{This figure shows the dependence of the nonthermal momentum distribution on $\mv$ and $\tau$. In all cases, $q$ and $f(q)$ are in units of the appropriate powers of the associated  $T_{\rm ncdm,0}$.}
 \label{taumv_fq}
 \end{figure}

Working with an extension of the 6-parameter $\Lambda$CDM model, here we include the mass ($m_{\rm SP}$) and momentum distribution $f(q)$ of an additional component of hot dark matter as inputs~\cite{class102}. We emphasise that in our implementation we explicitly use the non-thermal distribution function in \pref{fq}, this is done using the routine described in \cite{cl4}. The non-thermal momentum distribution $f(q)$ in our case depends on the mass $m_{\varphi}$ and lifetime $\tau$ of the decaying particle and therefore, we consider few benchmark points to arrive at $f(q)$. Here we note that reference  \cite{j2} suggested that for implementation in CLASS   thermal distributions should be used\footnote{Although \cite{j2} did not extract the predictions for the CMB and LSS using CLASS, it outlined a strategy for doing so.}, we disagree  as the precise from of the distribution function is absolutely necessary
for extracting the predictions from CLASS.

We can use the analytic expression obtained for  $\Delta N_{\rm eff}$ and $w_{\rm sp}$ to obtain
benchmark points for our inputs to CLASS. Note that \pref{ddnn}  implies that $\Delta N_{\rm eff}$ is essentially
determined by the branching ratio $B_{\rm sp}$. Given the bound $\Delta N_{\rm eff} \lesssim 0.3$, we  take
$B_{\rm sp} = 0.05$. This corresponds to $\Delta N_{\rm eff} =0.15$. Now, as described in the introduction, in our scenario it is natural to think of $\varp$ as the inflaton. Motivated by this,
we take $\mv = 10^{-6} M_{\rm pl}$.  We will take the lifetime of the $\varp$ as a phenomenological parameter, and consider
the points $\tau = { 10^{8} / \mv} $ and $\tau = {10^{9} /\mv} $. If the inflaton decays by a non-renormalizable interaction, then our choice for the lifetime corresponds decay via an interaction suppressed by approximately the GUT scale. In figure \ref{taumv_fq}, we
plot our distribution function for various values of $\mv$ and $\tau$.
\begin{figure}[h]
\centering
\includegraphics[width=0.49 \textwidth]{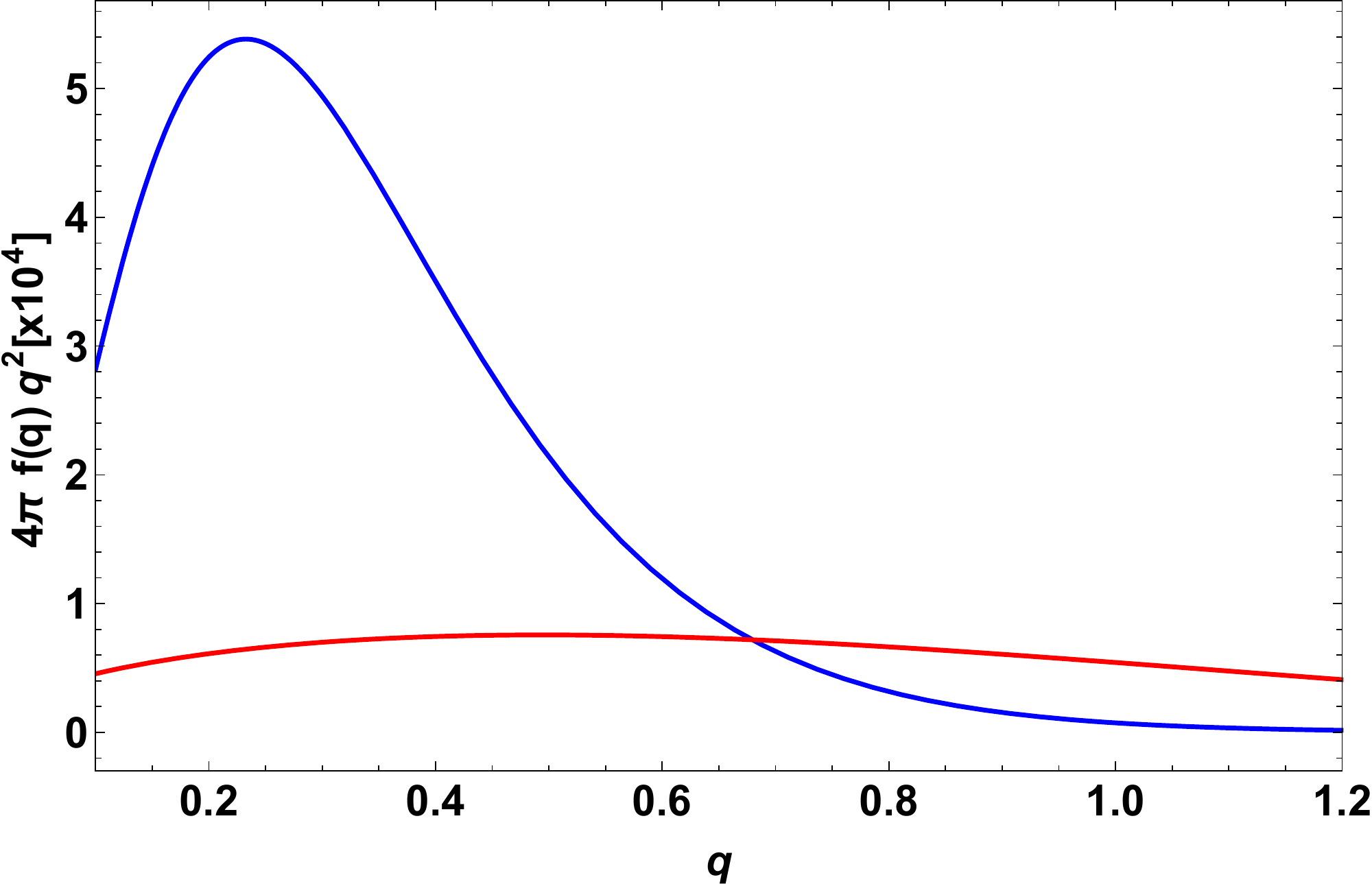}
\caption{Comparison with a thermal distribution with the  same value of $\Delta N_{\rm eff}$ (=0.15). The nonthermal distribution here is for $ \mv = 10^{-6} \mpl$, $\tau = {10^{8} / \mv}$ and is plotted in orange. The thermal 
distribution is in blue. The momenta  and the distribution functions for
both plots are in units of $T_{\rm ncdm,0}$ for the above value of $\mv$ and $\tau$.}
\label{dens}
\end{figure}  

It is interesting to compare our non-thermal distribution function with a thermal distribution function for sterile
neutrinos with the same value of $\Delta N_{\rm eff}$. We do this in figure \ref{dens}. Note that our distribution
function has a much lower maximum value but is much broader than the thermal one.
\begin{figure}
\includegraphics[width=0.5\textwidth]{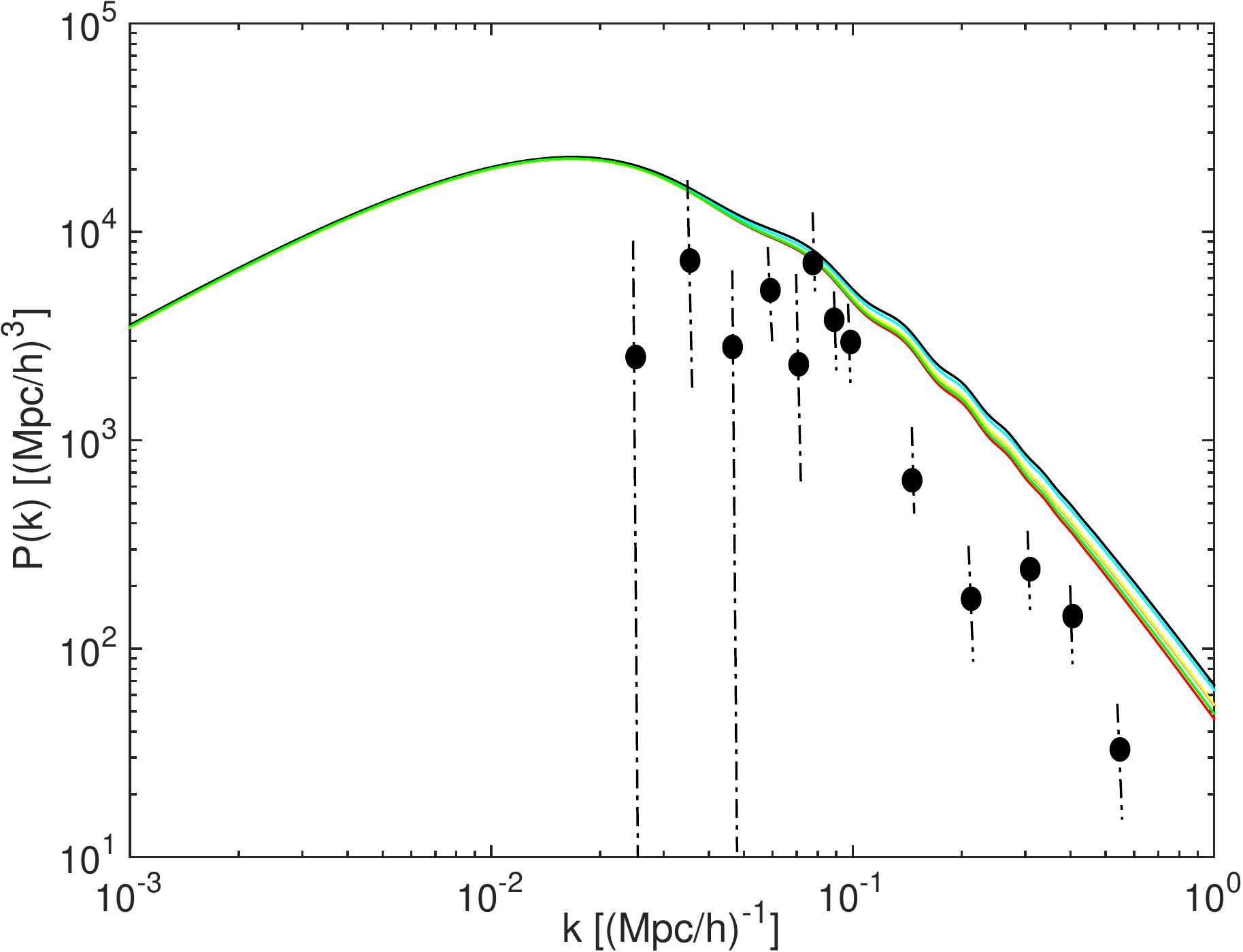}\\
\includegraphics[width=0.5\textwidth]{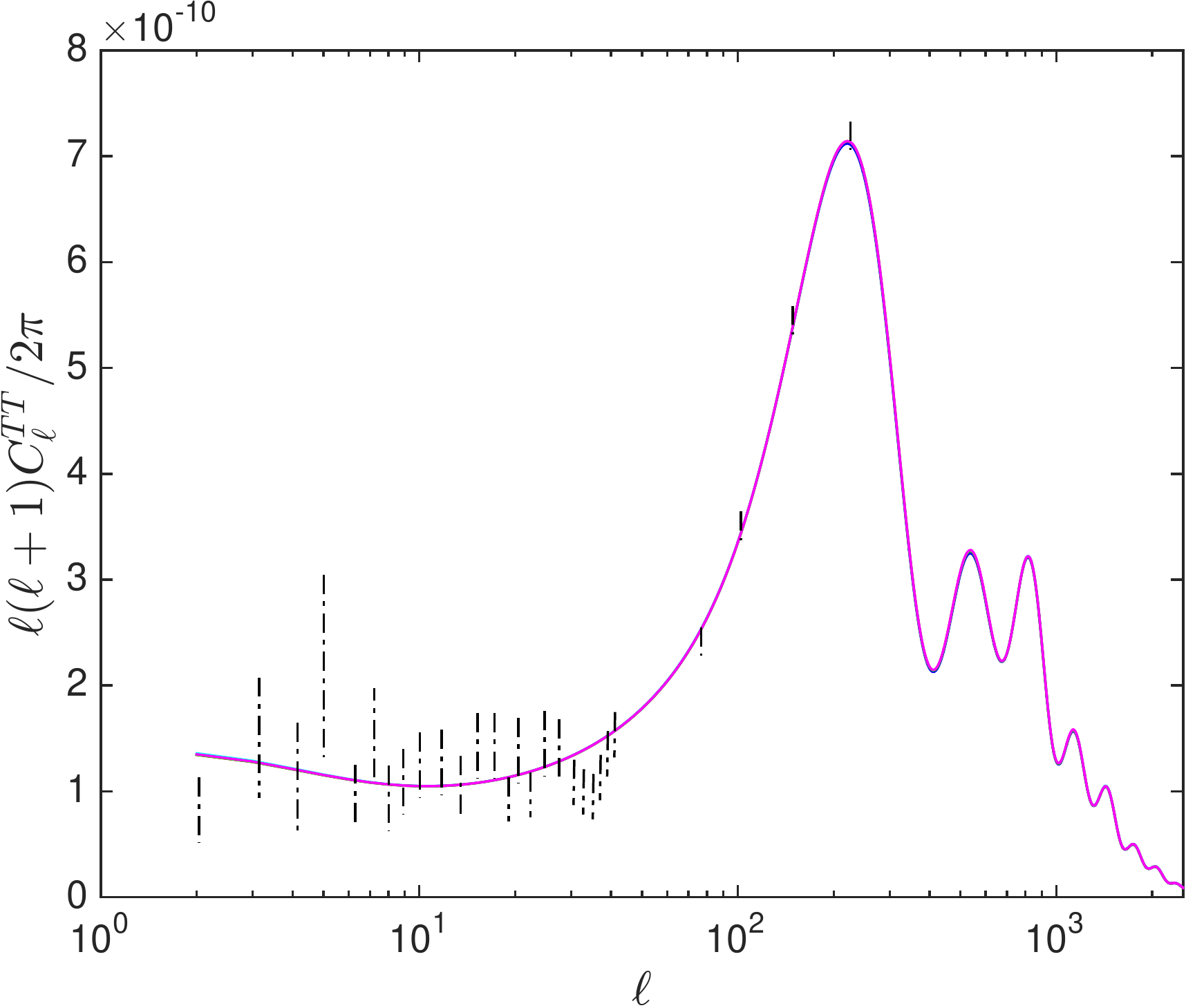}
\caption{Upper panel shows full matter power spectra $P(k)$ for $\tau=10^8/m_{\varphi}$ with inclusion of SN and SP of different masses. Colour schemes for different scenarios are the same as Fig.~\ref{fracClPktau89}. Central values and error-bars for the WiggleZ experiment are also given in black bold point and dot-dashed lines respectively on top of theoretical plots. The lower panel shows temperature power spectra $C_{\ell}^{TT}$ for the same benchmark point with error-bars from Planck 2018. $\theta _*$ is kept fixed in all the CLASS runs at the Planck 2018 TT+lowP value.}
  \label{Pktau89}
\end{figure}

\begin{figure*}[!t]
\subfigure{\includegraphics[width=0.48\textwidth]{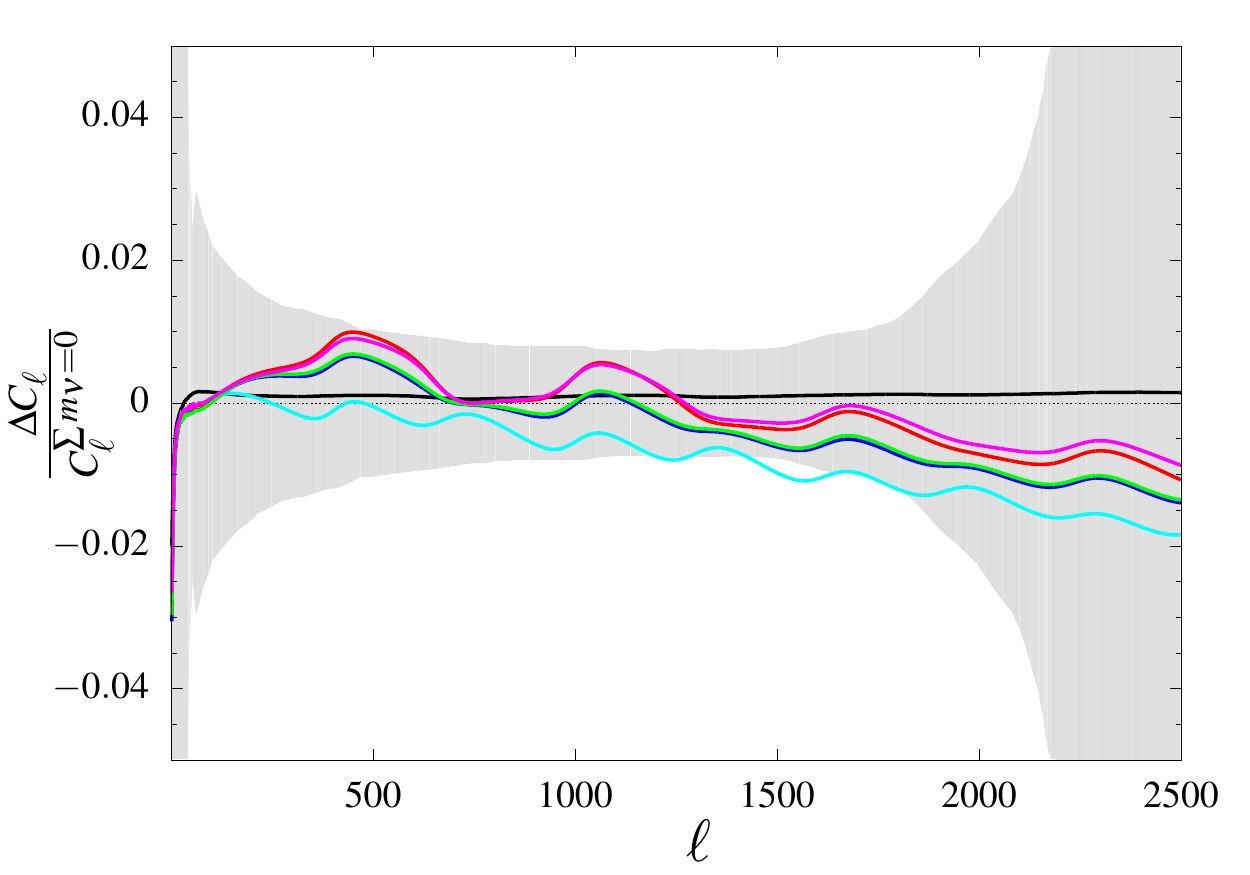}}
\subfigure{\includegraphics[width=0.48\textwidth]{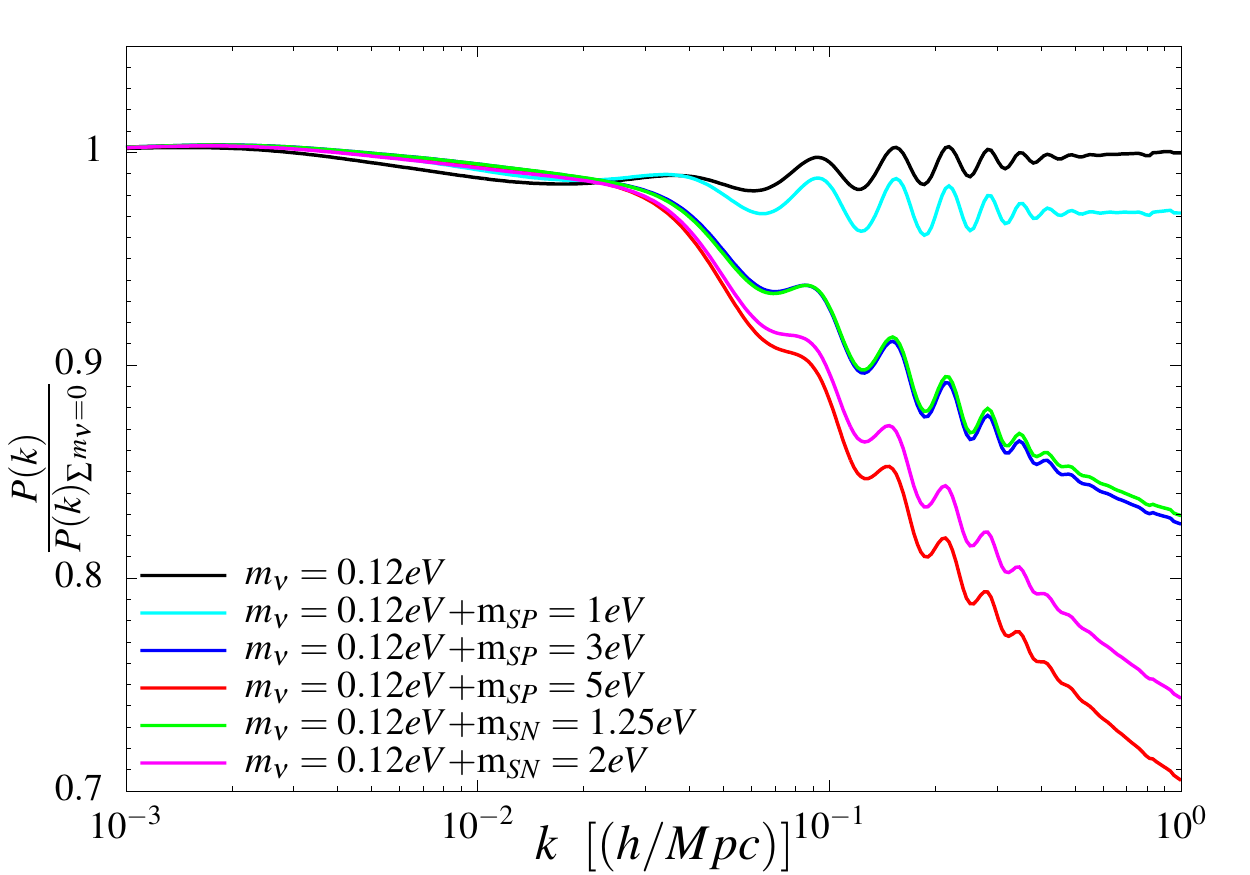}}
\subfigure{\includegraphics[width=0.48\textwidth]{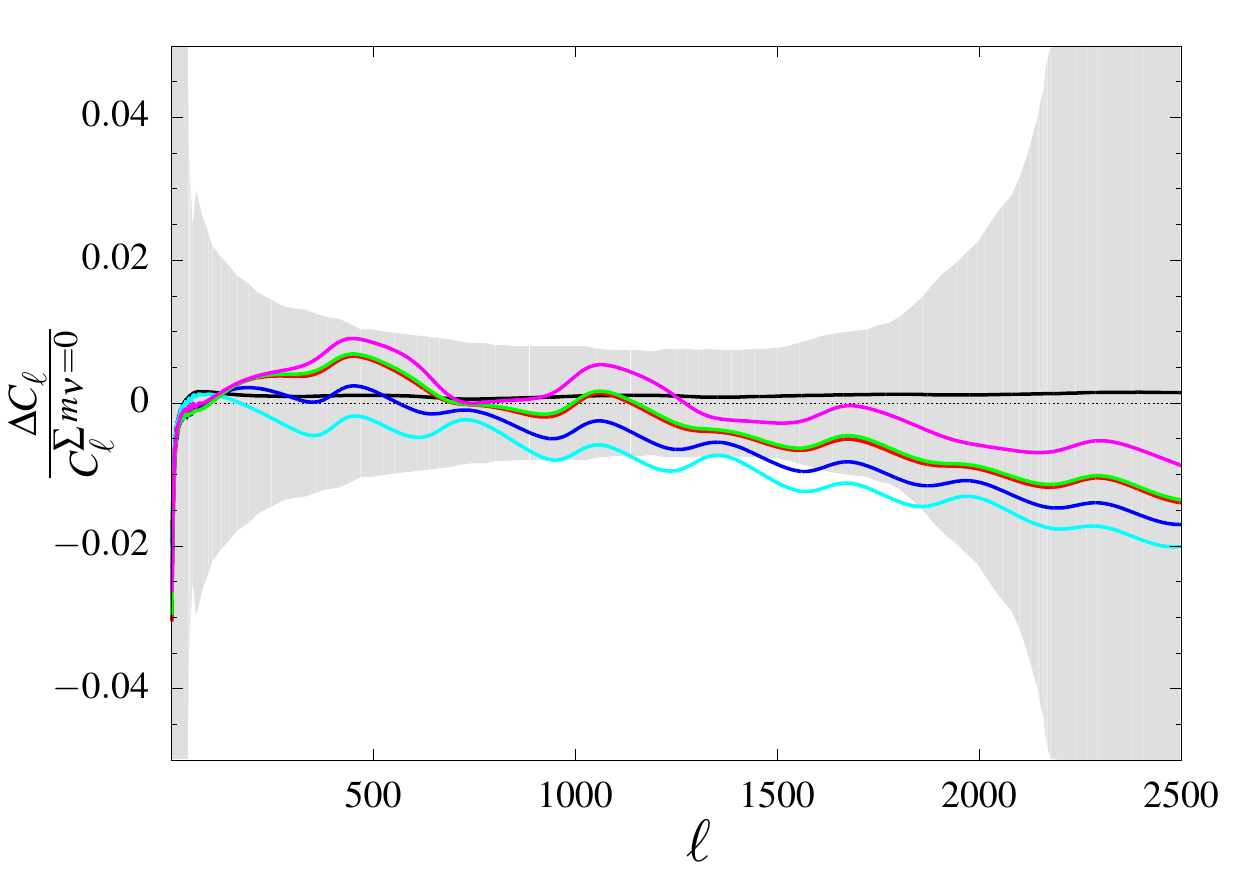}}
\subfigure{\includegraphics[width=0.48\textwidth]{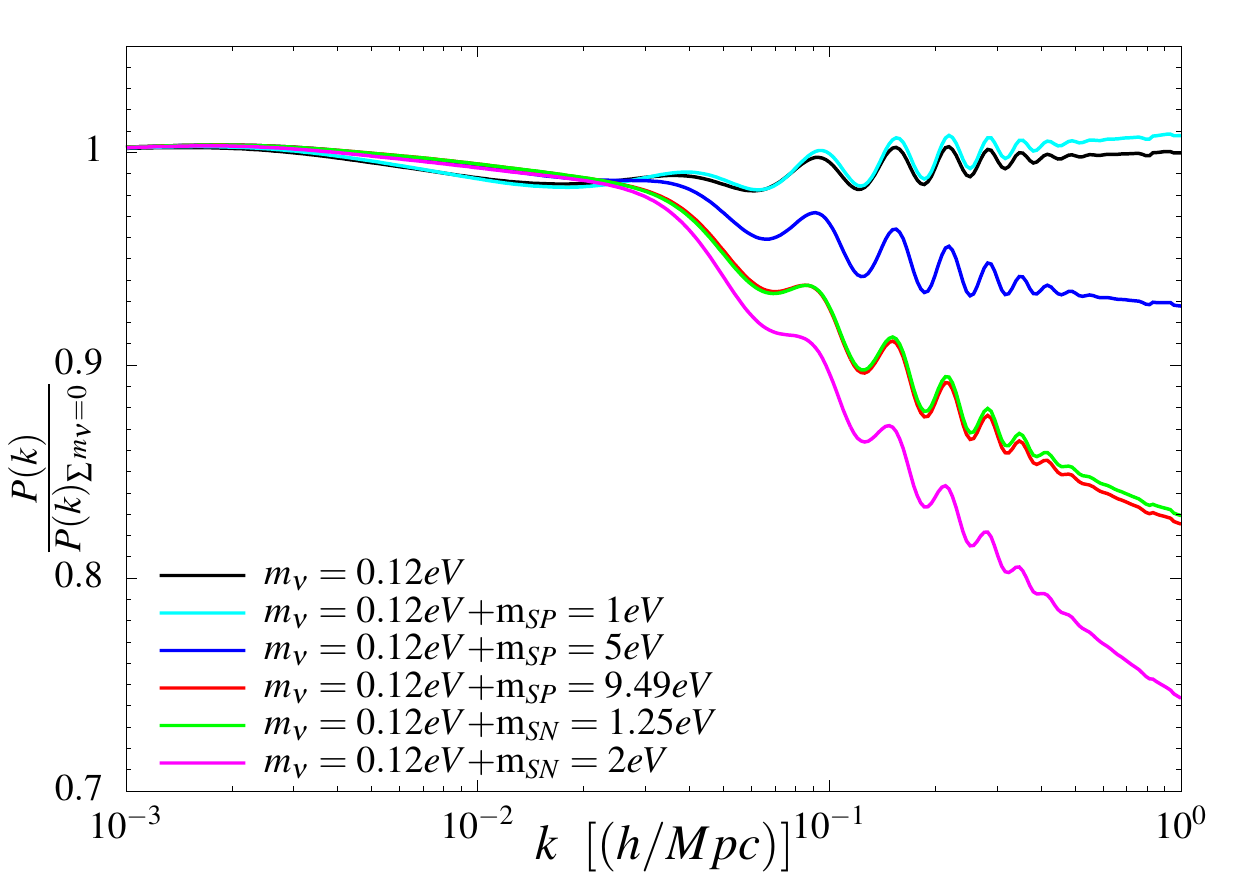}}
\caption{\small{Fractional deviation of the CMB temperature power spectra $C_{\ell}^{TT}$ (left panel) and fractional matter power spectra $P(k)$ (right panel) for different cases including sterile particles (SP) and sterile neutrino (SN) as shown in plot legends. For each case, $C_{\ell}^{TT}$ and $P(k)$ are evaluated for  $\sum m_{\nu}=0.12$ eV + SP/SN and compared to the case where $\sum m_{\nu}=0$ with no extra SP or SN species. The upper panels correspond to $\tau=10^8/m_{\varphi}$ and the lower panels are for $\tau=10^9/m_{\varphi}$. Note that for $\tau = 10^9 / m_{\varp} $, our hot dark matter
at $9.49 {\rm eV}$ is equivalent to a  $1.25  {\rm eV}$ thermalised neutrino (its temperature has been chosen such that both of them have the same $\Delta N_{\rm eff}$)}. The gray shaded region in the left panel corresponds to observed errors in $C_{\ell}^{TT}$ (Planck 2018).}
\label{fracClPktau89}
\end{figure*}

Taking these as inputs for CLASS we have computed 
 the matter power spectra $P(k)$ and temperature power spectra $C_{\ell}$ for the above discussed benchmark points
 for various values of $m_{\rm sp}$. In addition, for comparison we have plotted the effects of giving the standard model
 neutrinos a mass or sterile neutrinos (SN) at various masses. 
 The distribution for standard model neutrinos is taken to be thermal in the instantaneous decoupling  approximation $T_{\nu}^{\rm }=(4/13)^{1/3}T_{{\rm cmb},0}\simeq0.17$ meV. 
The SN are taken to be thermal with a temperature such that their contribution to $\Delta N_{\rm eff}$ is same
as that for our benchmark points (i.e $\Delta N_{\rm eff} = 0.15)$.
\comments{We have considered two masses of SN: for $m_{\rm SN}=1.25$ eV, we found that the temperature is $T_{\rm SN}\simeq 0.45 T_{{\rm cmb},0} \simeq 0.105$ meV, whereas for $m_{\rm SN}=2$ eV, $T_{\rm SN}\simeq 0.43 T_{{\rm cmb},0} \simeq 0.101$ meV.}
 For each of the benchmark points, the $T_{{\rm ncdm},0}$ is calculated from~\pref{tmomf}; i.e $T_{{\rm ncdm},0}=4.23T_{{\rm cmb},0}\simeq 1$ meV for $\tau=10^8/m_{\varphi}$ and $T_{{\rm ncdm},0}=13.39T_{{\rm cmb},0}\simeq 3.14$ meV for $\tau=10^9/m_{\varphi}$ which are then fed in the CLASS code. To keep the redshift $z_{\rm eq}$ of matter-radiation equality consistent at $z_{\rm eq}\simeq 3410$ for all these cases, the value of $\Omega_{\rm c}^{(0)}$ was modified slightly for each case. 
 The outcomes for $P(k)$ from CLASS are plotted in Figure~\ref{Pktau89}. The fractional changes in $P(k)$ and $C_{\ell}$ for such cases with respect to the case with no sterile particle with $\sum m_{\nu}=0$ are shown in Figure~\ref{fracClPktau89}. Now, we discuss two interesting aspects of our results.
\begin{itemize} 

\item As we can see from figure \ref{fracClPktau89}, the linear matter power spectra  gets much less suppression for our hot dark matter when compared to a standard thermalised neutrino of the same mass. The same is true for the effects in the CMB. For example, for $\tau = 10^9 / m_{\varp} $, we see that our hot dark matter
at $9.49 {\rm eV}$ is equivalent to a  $1.25  {\rm eV}$ thermalised neutrino (as discussed earlier, the 
temperature has been so chosen such that both of them have the same value of $\Delta N_{\rm eff}$). This matching can
be seen for our expression of $w_{\rm sp}$ in \pref{wff}. For the values corresponding to our benchmark point the effective
mass is seen to be reduced by one order of magnitude\footnote{Recall that the effective mass of a sterile species X
is defined by $m_{\rm eff}^{X} = \Omega_{X} h^2 94. 05 \phantom{a} {\rm eV} $.}. This brings to us an important point. For the thermal and Dodelson-Widrow
distributions \cite{Dodelson:1993je}, $\Delta N_{\rm eff}$  set the ratio of physical mass and the effective mass;
$ m_{\rm physical}^{\rm thermal} = (\Delta N_{\rm eff})^{-3/4}m_{\rm eff}^{\rm thermal}$ and
$m_{\rm physical}^{\rm DW} = (\Delta N_{\rm eff})^{-1}m_{\rm eff}^{\rm DW} $ (see e.g
\cite{pl18}). On the other hand, as seen from \pref{ddnn} $\Delta N_{\rm eff}$ is  set by the branching ratio, while $w_{\rm sp}$ has also
got dependence of the mass and lifetime of the decaying particle \pref{wff}. This makes $m_{\rm eff}$ and $\Delta N_{\rm eff}$
decoupled. This decoupling is essentially what allows for greater values of mass of our hot dark matter to be consistent with
the data. 

\item Another interesting feature is the $\ell$ dependence of on the effects on the CMB. 
 For CMB the main effect comes from $\Delta N_{eff}$ as it changes the Hubble expansion rate prior to photon decoupling \cite{1104.2333} which changes the silk damping scale. This effect shows up in higher $\ell$ (small scales) of CMB anisotropy power spectra which is evident from  figure \ref{fracClPktau89}. We see that lower the mass of our DM particle, higher is the effect in small scale as expected.  Where as there is another subtle effect when one introduces interacting dark radiation or non-thermal dark radiation.  Free-streaming hot DM travel supersonically through the photon-baryon plasma at early times, hence gravitationally pulling photon-baryon wave-fronts  slightly  ahead  of  where  they  would  be  in  the  absence  of  neutrinos.   As  a  result,  the free-streaming neutrinos imprint a net phase shift in the CMB power spectra towards larger scales (smaller $\ell$), as well as a suppression of its amplitude  \cite{Kreisch:2019yzn}.  In our case when we keep $\Delta N_{eff}$ more or less fixed given by Planck bound and as we vary hot DM mass, the distribution function also (as well as hot DM velocity) changes and the effects shows up in small $\ell$ values of CMB spectra.  We can find this effect in left panel of from figure \ref{fracClPktau89}, where in small $\ell$ values we see the deviations for different choices of mass of our hot DM particle. This, we find to be a very interesting effect and a detailed MCMC statistical analysis (which is work in progress) will make it clearer whether it can be detected by upcoming CMB and LSS experiments.
\end{itemize}

\section{Discussion and future directions}
\label{disc}
We have considered hot dark matter produced from decays in a universe transiting from
matter domination to radiation domination. Such epochs can occur naturally  during the perturbative
decay of the inflaton or as a result of vacuum misalignment of moduli fields. We have 
taken into account the characteristic moment distribution of   hot dark matter particles produced in this manner
and obtained their effects on the CMB and LSS making use of CLASS. Our analysis
has revealed interesting features such as higher values of hot dark matter
 mass being consistent with the linear matter power spectra and corresponding cosmological observations like Wiggle-Z. We have also found  features in the CMB at low $\ell$ potentially related to the phase difference appearing due to supersonic transmission of hot dark matter through the photon-baryon plasma before they turn non-relativistic.

 As mentioned above, a detailed MCMC statistical analysis to gain a better understanding
 of the cosmological implications is under progress. Other than this, there are many interesting avenues that can be pursued with the non-thermal distribution function. Cosmology has entered a high precision era not only with respect to the CMB but also through 
 non-linear  structure formation. It will be really interesting to  study structure formation  with a  non-thermal distribution function like ours. As has been pointed out in \cite{Bayer:2020tko}, the velocity  phase space distribution  plays a crucial role for non-linear structure formation in the presence of hot DM. Another future avenue is the study of  implications of the presence of such relativistic hot DM particles and the corresponding extra radiation like degrees of freedom for the Hubble anomaly \cite{DEramo:2018vss}. For a standard neutrino like particle, when one tries to address the Hubble anomaly by increasing the effective amount of radiation, it indeed tends to relax the tension but only partially \cite{Riess:2016jrr}. The reason for this is that an effective increase in the number of thermal  neutrino like particles makes the CMB high $\ell$ power deviate \cite{Kreisch:2019yzn} from the observed Planck value. It will be very interesting to see if this non-thermal distribution function could help us with the  high $\ell$ discrepancy. This is work in progress and will be reported in the near future.
 
 In this paper, we have not made an attempt to connect  to the  short base line anomaly \cite{AguilarArevalo:2010wv}, but it is worth pointing out that non-trivial momentum distribution functions as well as decay products might have implications for these anomalies. In our case, if the inflaton/moduli decay to intermediate mass sterile states which then decay into $eV$ sterile dark radiation, the idea presented in \cite{Dentler:2019dhz} can be relevant. We leave such a study for future work. 
 
  Our  set up can be easily extended for warm dark matter \cite{Abazajian:2017tcc} -- that is a scenario where the  mass of  the sterile particle is much higher (of the order of KeV).  The new distribution function will give rise to new results for WDM  from N-body simulations due to changes in the velocity phase space distribution\footnote{It will be very interesting to see if the effect of free-streaming is reduced or enhanced in comparison with similar WDM mass produced by standard mechanism such as  in \cite{Dodelson:1993je}.  Assuming WDM becomes non-relativistic in the radiation dominated epoch, in the approximation
  $\tau \gg t_{\rm nr}$ ($t_{\rm nr}$ being the time that the dark matter particle go non-relativistic), 
  free streaming length  is known to scale as  $t_{\rm nr} / a_{\rm nr}$. The results  in section \ref{distribute} give
 $$
   {t_{\rm nr} \over  a_{\rm nr}} \propto { \mv (M_{\rm pl} \tau)^{1/2} \over m_{\rm sp} T_{\rm cmb} }.
 $$
 So we see that depending on $ \tau$ and $m_{\rm sp}$ we can obtain various values of the free streaming length.}
\cite{Colin:2007bk} and new  constraints on  WDM mass from the  Lyman-alpha forest \cite{Boyarsky:2008xj}  and Milky Way (MW) satellites\cite{Nadler:2020prv,Das:2020nwc}. Again, work in this direction is in progress.

 
 Finally, there is lot of optimism that near future experiments  will be  able to distinguish or detect hot dark matter candidates with different particle physics origin (see e.g \cite{DePorzio:2020wcz}). For this, understanding the subtle effects which different hot DM particle imprint on CMB power spectra (for our case low $\ell$ phase shift, high $\ell$ suppression) that could  be measured by CMB-S4 \cite{abazajian2019cmbs4} experiments is very important. The same is true for ongoing or upcoming LSS experiments like BOSS \cite{Alam_2017}, DESI \cite{collaboration2016desi},
 EUCLID \cite{Amendola_2018} DES\cite{1708.01530} and KiDS\cite{ 1812.06076} which will measure the  linear matter power spectra with high accuracy and may be able to distinguish between thermal and non-thermal suppression. It will be very interesting see if our model can also relax recent $\sigma_8$ anomaly between CMB and weak lensing results. The present work together with the ongoing MCMC analysis should provide an interesting theory input for all of this.

\section*{Acknowledgments}
We thank Vivian Poulin and Arka Banerjee for comments on the manuscript and Shiv Sethi, Shouvik Roy Choudhury and Raj Gandhi for useful suggestions. 
SB is supported by postdoctoral fellowship from Physical Research Laboratory, India. Work of MRG is supported by the Department of Science and Technology, Government of India under the Grant Agreement number IF18-PH-228 (INSPIRE Faculty Award). AM is supported
in part by the SERB, DST, Government of India by the grant MTR/2019/000267. KD is supported in part by the grant MTR/2019/000395, funded by the DST, Govt of India. AM would like to thank the Department of Physics,
National Taiwan University for hospitality. SD acknowledges SERB grant CRG/2019/006147. SB and MRG acknowledge the hospitality of the Harish-Chandra Research Institute during the early stages of this project.



\bibliographystyle{utphys}

\bibliography{mybib}

\providecommand{\href}[2]{#2}\begingroup\raggedright\begin{thebibliography}{100}

\bibitem{Bertone:2010zza}
J.~Silk {\em et~al.}, \href{http://dx.doi.org/10.1017/CBO9780511770739}{{\em
  {Particle Dark Matter: Observations, Models and Searches}}}.
\newblock Cambridge Univ. Press, Cambridge, 2010.

\bibitem{Mukhanov:2005sc}
V.~Mukhanov, {\em {Physical Foundations of Cosmology}}.
\newblock Cambridge University Press, Oxford, 2005.

\bibitem{mod1}
G.~Coughlan, W.~Fischler, E.~W. Kolb, S.~Raby, and G.~G. Ross, ``{Cosmological
  Problems for the Polonyi Potential},''
  \href{http://dx.doi.org/10.1016/0370-2693(83)91091-2}{{\em Phys. Lett. B}
  {\bfseries 131} (1983) 59--64}.

\bibitem{fq}
B.~de~Carlos, J.~Casas, F.~Quevedo, and E.~Roulet, ``{Model independent
  properties and cosmological implications of the dilaton and moduli sectors of
  4-d strings},'' \href{http://dx.doi.org/10.1016/0370-2693(93)91538-X}{{\em
  Phys. Lett. B} {\bfseries 318} (1993) 447--456},
  \href{http://arxiv.org/abs/hep-ph/9308325}{{\ttfamily arXiv:hep-ph/9308325}}.

\bibitem{banks}
T.~Banks, D.~B. Kaplan, and A.~E. Nelson, ``{Cosmological implications of
  dynamical supersymmetry breaking},''
  \href{http://dx.doi.org/10.1103/PhysRevD.49.779}{{\em Phys. Rev. D}
  {\bfseries 49} (1994) 779--787},
  \href{http://arxiv.org/abs/hep-ph/9308292}{{\ttfamily arXiv:hep-ph/9308292}}.

\bibitem{Cicoli:2016olq}
M.~Cicoli, K.~Dutta, A.~Maharana, and F.~Quevedo, ``{Moduli Vacuum Misalignment
  and Precise Predictions in String Inflation},''
  \href{http://dx.doi.org/10.1088/1475-7516/2016/08/006}{{\em JCAP} {\bfseries
  08} (2016) 006}, \href{http://arxiv.org/abs/1604.08512}{{\ttfamily
  arXiv:1604.08512 [hep-th]}}.

\bibitem{Kane:2015jia}
G.~Kane, K.~Sinha, and S.~Watson, ``{Cosmological Moduli and the
  Post-Inflationary Universe: A Critical Review},''
  \href{http://dx.doi.org/10.1142/S0218271815300220}{{\em Int. J. Mod. Phys. D}
  {\bfseries 24} no.~08, (2015) 1530022},
  \href{http://arxiv.org/abs/1502.07746}{{\ttfamily arXiv:1502.07746
  [hep-th]}}.

\bibitem{amin}
R.~Allahverdi {\em et~al.}, ``{The First Three Seconds: a Review of Possible
  Expansion Histories of the Early Universe},''
  \href{http://arxiv.org/abs/2006.16182}{{\ttfamily arXiv:2006.16182
  [astro-ph.CO]}}.

\bibitem{jim}
J.~Halverson and P.~Langacker, ``{TASI Lectures on Remnants from the String
  Landscape},'' \href{http://dx.doi.org/10.22323/1.305.0019}{{\em PoS}
  {\bfseries TASI2017} (2018) 019},
  \href{http://arxiv.org/abs/1801.03503}{{\ttfamily arXiv:1801.03503
  [hep-th]}}.

\bibitem{oturner}
R.~J. Scherrer and M.~S. Turner, ``{Primordial Nucleosynthesis with Decaying
  Particles. 1. Entropy Producing Decays. 2. Inert Decays},''
  \href{http://dx.doi.org/10.1086/166534}{{\em Astrophys. J.} {\bfseries 331}
  (1988) 19--32}.

\bibitem{cm}
J.~P. Conlon and M.~C.~D. Marsh, ``{The Cosmophenomenology of Axionic Dark
  Radiation},'' \href{http://dx.doi.org/10.1007/JHEP10(2013)214}{{\em JHEP}
  {\bfseries 10} (2013) 214}, \href{http://arxiv.org/abs/1304.1804}{{\ttfamily
  arXiv:1304.1804 [hep-ph]}}.

\bibitem{class101}
J.~{Lesgourgues}, ``{The Cosmic Linear Anisotropy Solving System (CLASS) I:
  Overview},'' {\em arXiv e-prints} (Apr., 2011) arXiv:1104.2932,
  \href{http://arxiv.org/abs/1104.2932}{{\ttfamily arXiv:1104.2932
  [astro-ph.IM]}}.

\bibitem{class102}
J.~{Lesgourgues} and T.~{Tram}, ``{The Cosmic Linear Anisotropy Solving System
  (CLASS) IV: efficient implementation of non-cold relics},''
  \href{http://dx.doi.org/10.1088/1475-7516/2011/09/032}{{\em JCAP} {\bfseries
  2011} no.~9, (Sept., 2011) 032},
  \href{http://arxiv.org/abs/1104.2935}{{\ttfamily arXiv:1104.2935
  [astro-ph.CO]}}.

\bibitem{cl4}
J.~Lesgourgues and T.~Tram, ``{The Cosmic Linear Anisotropy Solving System
  (CLASS) IV: efficient implementation of non-cold relics},''
  \href{http://dx.doi.org/10.1088/1475-7516/2011/09/032}{{\em JCAP} {\bfseries
  09} (2011) 032}, \href{http://arxiv.org/abs/1104.2935}{{\ttfamily
  arXiv:1104.2935 [astro-ph.CO]}}.

\bibitem{Dodelson:1993je}
S.~Dodelson and L.~M. Widrow, ``Sterile neutrinos as dark matter,''
  \href{http://dx.doi.org/10.1103/physrevlett.72.17}{{\em Physical Review
  Letters} {\bfseries 72} no.~1, (Jan., 1994) 17--20}.
  \url{https://doi.org/10.1103/physrevlett.72.17}.

\bibitem{Valdarnini:1998zy}
R.~Valdarnini, T.~Kahniashvili, and B.~Novosyadlyj, ``Large scale structure
  formation in mixed dark matter models with a cosmological constant,'' {\em
  Astronomy and Astrophysics} {\bfseries 336} (05, 1998) 11--28,
  \href{http://arxiv.org/abs/astro-ph/9804057v1}{{\ttfamily
  arXiv:astro-ph/9804057v1 [astro-ph]}}.

\bibitem{Boyarsky:2008mt}
A.~Boyarsky, J.~Lesgourgues, O.~Ruchayskiy, and M.~Viel, ``Realistic sterile
  neutrino dark matter with {keV} mass does not contradict cosmological
  bounds,'' \href{http://dx.doi.org/10.1103/physrevlett.102.201304}{{\em
  Physical Review Letters} {\bfseries 102} no.~20, (May, 2009) }.
  \url{https://doi.org/10.1103/physrevlett.102.201304}.

\bibitem{Bezrukov:2009th}
F.~Bezrukov, H.~Hettmansperger, and M.~Lindner, ``{keV} sterile neutrino dark
  matter in gauge extensions of the standard model,''
  \href{http://dx.doi.org/10.1103/physrevd.81.085032}{{\em Physical Review D}
  {\bfseries 81} no.~8, (Apr., 2010) }.
  \url{https://doi.org/10.1103/physrevd.81.085032}.

\bibitem{Petraki:2007gq}
K.~Petraki and A.~Kusenko, ``Dark-matter sterile neutrinos in models with a
  gauge singlet in the {Higgs} sector,''
  \href{http://dx.doi.org/10.1103/physrevd.77.065014}{{\em Physical Review D}
  {\bfseries 77} no.~6, (Mar., 2008) }.
  \url{https://doi.org/10.1103/physrevd.77.065014}.

\bibitem{Laine:2008pg}
M.~Laine and M.~Shaposhnikov, ``Sterile neutrino dark matter as a consequence
  of $\nu${MSM}-induced lepton asymmetry,''
  \href{http://dx.doi.org/10.1088/1475-7516/2008/06/031}{{\em Journal of
  Cosmology and Astroparticle Physics} {\bfseries 2008} no.~06, (June, 2008)
  031}. \url{https://doi.org/10.1088/1475-7516/2008/06/031}.

\bibitem{Abazajian:2005xn}
K.~Abazajian, ``Linear cosmological structure limits on warm dark matter,''
  \href{http://dx.doi.org/10.1103/physrevd.73.063513}{{\em Physical Review D}
  {\bfseries 73} no.~6, (Mar., 2006) }.
  \url{https://doi.org/10.1103/physrevd.73.063513}.

\bibitem{Irsic:2017ixq}
V.~Ir{\v{s}}i{\v{c}}~{\it et al.}, ``New constraints on the free-streaming of
  warm dark matter from intermediate and small scale lyman-$\alpha$ forest
  data,'' \href{http://dx.doi.org/10.1103/physrevd.96.023522}{{\em Physical
  Review D} {\bfseries 96} no.~2, (July, 2017) }.
  \url{https://doi.org/10.1103/physrevd.96.023522}.

\bibitem{Abazajian:2019ejt}
K.~N. Abazajian and A.~Kusenko, ``{Hidden treasures: Sterile neutrinos as dark
  matter with miraculous abundance, structure formation for different
  production mechanisms, and a solution to the $\sigma_8$ problem},''
  \href{http://dx.doi.org/10.1103/PhysRevD.100.103513}{{\em Phys. Rev. D}
  {\bfseries 100} no.~10, (2019) 103513},
  \href{http://arxiv.org/abs/1907.11696}{{\ttfamily arXiv:1907.11696
  [hep-ph]}}.

\bibitem{Samanta:2020gdw}
R.~Samanta, A.~Biswas, and S.~Bhattacharya, ``{Non-thermal production of lepton
  asymmetry and dark matter in minimal seesaw with right handed neutrino
  induced Higgs potential},'' \href{http://arxiv.org/abs/2006.02960}{{\ttfamily
  arXiv:2006.02960 [hep-ph]}}.

\bibitem{pl18}
{\bfseries Planck} Collaboration, N.~Aghanim {\em et~al.}, ``{Planck 2018
  results. VI. Cosmological parameters},''
  \href{http://arxiv.org/abs/1807.06209}{{\ttfamily arXiv:1807.06209
  [astro-ph.CO]}}.

\bibitem{Das:2020wfe}
S.~Das and P.~Chanda, ``{Degenerate dark matter micro-nuggets from $\rm{eV}$
  sterile states and the Hubble tension},''
  \href{http://arxiv.org/abs/2005.11889}{{\ttfamily arXiv:2005.11889
  [astro-ph.CO]}}.

\bibitem{Das:2006ht}
S.~Das and N.~Weiner, ``{Late Forming Dark Matter in Theories of Neutrino Dark
  Energy},'' \href{http://dx.doi.org/10.1103/PhysRevD.84.123511}{{\em Phys.
  Rev. D} {\bfseries 84} (2011) 123511},
  \href{http://arxiv.org/abs/astro-ph/0611353}{{\ttfamily
  arXiv:astro-ph/0611353}}.

\bibitem{Aguilar-Arevalo:2018gpe}
A.~A. Aguilar-Arevalo~{\it et al.}, ``Significant excess of electronlike events
  in the {MiniBooNE} short-baseline neutrino experiment,''
  \href{http://dx.doi.org/10.1103/physrevlett.121.221801}{{\em Physical Review
  Letters} {\bfseries 121} no.~22, (Nov., 2018) }.
  \url{https://doi.org/10.1103/physrevlett.121.221801}.

\bibitem{Aartsen:2017bap}
{\bfseries IceCube} Collaboration, M.~Aartsen {\em et~al.}, ``{Search for
  sterile neutrino mixing using three years of IceCube DeepCore data},''
  \href{http://dx.doi.org/10.1103/PhysRevD.95.112002}{{\em Phys. Rev. D}
  {\bfseries 95} no.~11, (2017) 112002},
  \href{http://arxiv.org/abs/1702.05160}{{\ttfamily arXiv:1702.05160
  [hep-ex]}}.

\bibitem{Adamson:2017zcg}
{\bfseries NOvA} Collaboration, P.~Adamson {\em et~al.}, ``{Search for
  active-sterile neutrino mixing using neutral-current interactions in NOvA},''
  \href{http://dx.doi.org/10.1103/PhysRevD.96.072006}{{\em Phys. Rev. D}
  {\bfseries 96} no.~7, (2017) 072006},
  \href{http://arxiv.org/abs/1706.04592}{{\ttfamily arXiv:1706.04592
  [hep-ex]}}.

\bibitem{Kreisch:2019yzn}
C.~D. Kreisch, F.-Y. Cyr-Racine, and O.~Doré, ``{The Neutrino Puzzle:
  Anomalies, Interactions, and Cosmological Tensions},''
  \href{http://dx.doi.org/10.1103/PhysRevD.101.123505}{{\em Phys. Rev. D}
  {\bfseries 101} no.~12, (2020) 123505},
  \href{http://arxiv.org/abs/1902.00534}{{\ttfamily arXiv:1902.00534
  [astro-ph.CO]}}.

\bibitem{Dasgupta:2013zpn}
B.~Dasgupta and J.~Kopp, ``Cosmologically safe {eV}-scale sterile neutrinos and
  improved dark matter structure,''
  \href{http://dx.doi.org/10.1103/physrevlett.112.031803}{{\em Physical Review
  Letters} {\bfseries 112} no.~3, (Jan., 2014) }.
  \url{https://doi.org/10.1103/physrevlett.112.031803}.

\bibitem{Dentler:2019dhz}
M.~Dentler, I.~Esteban, J.~Kopp, and P.~Machado, ``{Decaying Sterile Neutrinos
  and the Short Baseline Oscillation Anomalies},''
  \href{http://dx.doi.org/10.1103/PhysRevD.101.115013}{{\em Phys. Rev. D}
  {\bfseries 101} no.~11, (2020) 115013},
  \href{http://arxiv.org/abs/1911.01427}{{\ttfamily arXiv:1911.01427
  [hep-ph]}}.

\bibitem{Dodelson:2005tp}
S.~Dodelson, A.~Melchiorri, and A.~Slosar, ``{Is cosmology compatible with
  sterile neutrinos?},''
  \href{http://dx.doi.org/10.1103/PhysRevLett.97.041301}{{\em Phys. Rev. Lett.}
  {\bfseries 97} (2006) 041301},
  \href{http://arxiv.org/abs/astro-ph/0511500}{{\ttfamily
  arXiv:astro-ph/0511500}}.

\bibitem{s2}
M.~A. Acero and J.~Lesgourgues, ``{Cosmological constraints on a light
  non-thermal sterile neutrino},''
  \href{http://dx.doi.org/10.1103/PhysRevD.79.045026}{{\em Phys. Rev. D}
  {\bfseries 79} (2009) 045026},
  \href{http://arxiv.org/abs/0812.2249}{{\ttfamily arXiv:0812.2249
  [astro-ph]}}.

\bibitem{Hannestad:2012ky}
S.~Hannestad, I.~Tamborra, and T.~Tram, ``{Thermalisation of light sterile
  neutrinos in the early universe},''
  \href{http://dx.doi.org/10.1088/1475-7516/2012/07/025}{{\em JCAP} {\bfseries
  07} (2012) 025}, \href{http://arxiv.org/abs/1204.5861}{{\ttfamily
  arXiv:1204.5861 [astro-ph.CO]}}.

\bibitem{Oldengott:2019lke}
I.~M. Oldengott, G.~Barenboim, S.~Kahlen, J.~Salvado, and D.~J. Schwarz, ``{How
  to relax the cosmological neutrino mass bound},''
  \href{http://dx.doi.org/10.1088/1475-7516/2019/04/049}{{\em JCAP} {\bfseries
  04} (2019) 049}, \href{http://arxiv.org/abs/1901.04352}{{\ttfamily
  arXiv:1901.04352 [astro-ph.CO]}}.

\bibitem{1104.2333}
Z.~Hou, R.~Keisler, L.~Knox, M.~Millea, and C.~Reichardt, ``How massless
  neutrinos affect the cosmic microwave background damping tail,''
  \href{http://dx.doi.org/10.1103/physrevd.87.083008}{{\em Physical Review D}
  {\bfseries 87} no.~8, (Apr, 2013) }.
  \url{http://dx.doi.org/10.1103/PhysRevD.87.083008}.

\bibitem{jj}
J.~Hasenkamp and J.~Kersten, ``{Dark radiation from particle decay:
  cosmological constraints and opportunities},''
  \href{http://dx.doi.org/10.1088/1475-7516/2013/08/024}{{\em JCAP} {\bfseries
  08} (2013) 024}, \href{http://arxiv.org/abs/1212.4160}{{\ttfamily
  arXiv:1212.4160 [hep-ph]}}.

\bibitem{j2}
J.~Hasenkamp, ``{Daughters mimic sterile neutrinos (almost!) perfectly},''
  \href{http://dx.doi.org/10.1088/1475-7516/2014/09/048}{{\em JCAP} {\bfseries
  09} (2014) 048}, \href{http://arxiv.org/abs/1405.6736}{{\ttfamily
  arXiv:1405.6736 [astro-ph.CO]}}.

\bibitem{Choudhury:2018sbz}
S.~Roy~Choudhury and S.~Choubey, ``{Constraining light sterile neutrino mass
  with the BICEP2/Keck Array 2014 B-mode polarization data},''
  \href{http://dx.doi.org/10.1140/epjc/s10052-019-7063-2}{{\em Eur. Phys. J. C}
  {\bfseries 79} no.~7, (2019) 557},
  \href{http://arxiv.org/abs/1807.10294}{{\ttfamily arXiv:1807.10294
  [astro-ph.CO]}}.

\bibitem{astro-ph/0404585}
J.~F. Beacom, N.~F. Bell, and S.~Dodelson, ``{Neutrinoless universe},''
  \href{http://dx.doi.org/10.1103/PhysRevLett.93.121302}{{\em Phys. Rev. Lett.}
  {\bfseries 93} (2004) 121302},
\href{http://arxiv.org/abs/astro-ph/0404585}{{\ttfamily arXiv:astro-ph/0404585
  [astro-ph]}}.

\bibitem{hep-ph/0402049}
P.~Crotty, J.~Lesgourgues, and S.~Pastor, ``{Current cosmological bounds on
  neutrino masses and relativistic relics},''
  \href{http://dx.doi.org/10.1103/PhysRevD.69.123007}{{\em Phys. Rev.}
  {\bfseries D69} (2004) 123007},
\href{http://arxiv.org/abs/hep-ph/0402049}{{\ttfamily arXiv:hep-ph/0402049
  [hep-ph]}}.

\bibitem{astro-ph/0502465}
A.~Cuoco, J.~Lesgourgues, G.~Mangano, and S.~Pastor, ``{Do observations prove
  that cosmological neutrinos are thermally distributed?},''
  \href{http://dx.doi.org/10.1103/PhysRevD.71.123501}{{\em Phys. Rev.}
  {\bfseries D71} (2005) 123501},
\href{http://arxiv.org/abs/astro-ph/0502465}{{\ttfamily arXiv:astro-ph/0502465
  [astro-ph]}}.

\bibitem{astro-ph/0607086}
M.~Cirelli and A.~Strumia, ``{Cosmology of neutrinos and extra light particles
  after WMAP3},'' \href{http://dx.doi.org/10.1088/1475-7516/2006/12/013}{{\em
  JCAP} {\bfseries 0612} (2006) 013},
\href{http://arxiv.org/abs/astro-ph/0607086}{{\ttfamily arXiv:astro-ph/0607086
  [astro-ph]}}.

\bibitem{hep-ph/0012317}
G.~F. Giudice, E.~W. Kolb, A.~Riotto, D.~V. Semikoz, and I.~I. Tkachev,
  ``{Standard model neutrinos as warm dark matter},''
  \href{http://dx.doi.org/10.1103/PhysRevD.64.043512}{{\em Phys. Rev.}
  {\bfseries D64} (2001) 043512},
\href{http://arxiv.org/abs/hep-ph/0012317}{{\ttfamily arXiv:hep-ph/0012317
  [hep-ph]}}.

\bibitem{astro-ph/0403323}
G.~Gelmini, S.~Palomares-Ruiz, and S.~Pascoli, ``{Low reheating temperature and
  the visible sterile neutrino},''
  \href{http://dx.doi.org/10.1103/PhysRevLett.93.081302}{{\em Phys. Rev. Lett.}
  {\bfseries 93} (2004) 081302},
\href{http://arxiv.org/abs/astro-ph/0403323}{{\ttfamily arXiv:astro-ph/0403323
  [astro-ph]}}.

\bibitem{hep-ph/9303287}
S.~Dodelson and L.~M. Widrow, ``{Sterile-neutrinos as dark matter},''
  \href{http://dx.doi.org/10.1103/PhysRevLett.72.17}{{\em Phys. Rev. Lett.}
  {\bfseries 72} (1994) 17--20},
\href{http://arxiv.org/abs/hep-ph/9303287}{{\ttfamily arXiv:hep-ph/9303287
  [hep-ph]}}.

\bibitem{astro-ph/9505029}
S.~Colombi, S.~Dodelson, and L.~M. Widrow, ``{Large scale structure tests of
  warm dark matter},'' \href{http://dx.doi.org/10.1086/176788}{{\em Astrophys.
  J.} {\bfseries 458} (1996) 1},
\href{http://arxiv.org/abs/astro-ph/9505029}{{\ttfamily arXiv:astro-ph/9505029
  [astro-ph]}}.

\bibitem{0803.2735}
G.~Gelmini, E.~Osoba, S.~Palomares-Ruiz, and S.~Pascoli, ``{MeV sterile
  neutrinos in low reheating temperature cosmological scenarios},''
  \href{http://dx.doi.org/10.1088/1475-7516/2008/10/029}{{\em JCAP} {\bfseries
  0810} (2008) 029},
\href{http://arxiv.org/abs/0803.2735}{{\ttfamily arXiv:0803.2735 [astro-ph]}}.

\bibitem{hep-ph/0504059}
S.~Hannestad, A.~Mirizzi, and G.~Raffelt, ``{New cosmological mass limit on
  thermal relic axions},''
  \href{http://dx.doi.org/10.1088/1475-7516/2005/07/002}{{\em JCAP} {\bfseries
  0507} (2005) 002},
\href{http://arxiv.org/abs/hep-ph/0504059}{{\ttfamily arXiv:hep-ph/0504059
  [hep-ph]}}.

\bibitem{0803.1585}
S.~Hannestad, A.~Mirizzi, G.~G. Raffelt, and Y.~Y.~Y. Wong, ``{Cosmological
  constraints on neutrino plus axion hot dark matter: Update after WMAP-5},''
  \href{http://dx.doi.org/10.1088/1475-7516/2008/04/019}{{\em JCAP} {\bfseries
  0804} (2008) 019},
\href{http://arxiv.org/abs/0803.1585}{{\ttfamily arXiv:0803.1585 [astro-ph]}}.

\bibitem{astro-ph/0302337}
P.~Crotty, J.~Lesgourgues, and S.~Pastor, ``{Measuring the cosmological
  background of relativistic particles with WMAP},''
  \href{http://dx.doi.org/10.1103/PhysRevD.67.123005}{{\em Phys. Rev.}
  {\bfseries D67} (2003) 123005},
\href{http://arxiv.org/abs/astro-ph/0302337}{{\ttfamily arXiv:astro-ph/0302337
  [astro-ph]}}.

\bibitem{hep-ph/0412181}
S.~Hannestad, ``{Neutrino mass bounds from cosmology},''
  \href{http://dx.doi.org/10.1016/j.nuclphysbps.2005.04.029}{{\em Nucl. Phys.
  Proc. Suppl.} {\bfseries 145} (2005) 313--318},
\href{http://arxiv.org/abs/hep-ph/0412181}{{\ttfamily arXiv:hep-ph/0412181
  [hep-ph]}}.

\bibitem{astro-ph/0105385}
S.~H. Hansen, G.~Mangano, A.~Melchiorri, G.~Miele, and O.~Pisanti,
  ``{Constraining neutrino physics with BBN and CMBR},''
  \href{http://dx.doi.org/10.1103/PhysRevD.65.023511}{{\em Phys. Rev.}
  {\bfseries D65} (2002) 023511},
\href{http://arxiv.org/abs/astro-ph/0105385}{{\ttfamily arXiv:astro-ph/0105385
  [astro-ph]}}.

\bibitem{astro-ph/0412066}
R.~Trotta and A.~Melchiorri, ``{Indication for primordial anisotropies in the
  neutrino background from WMAP and SDSS},''
  \href{http://dx.doi.org/10.1103/PhysRevLett.95.011305}{{\em Phys. Rev. Lett.}
  {\bfseries 95} (2005) 011305},
\href{http://arxiv.org/abs/astro-ph/0412066}{{\ttfamily arXiv:astro-ph/0412066
  [astro-ph]}}.

\bibitem{astro-ph/0503612}
A.~D. Dolgov, S.~H. Hansen, and A.~{\relax Yu}. Smirnov, ``{Neutrino statistics
  and Big Bang nucleosynthesis},''
  \href{http://dx.doi.org/10.1088/1475-7516/2005/06/004}{{\em JCAP} {\bfseries
  0506} (2005) 004},
\href{http://arxiv.org/abs/astro-ph/0503612}{{\ttfamily arXiv:astro-ph/0503612
  [astro-ph]}}.

\bibitem{astro-ph/0309135}
P.~Adhya, D.~R. Chaudhuri, and S.~Hannestad, ``{Late-time entropy production
  from scalar decay and relic neutrino temperature},''
  \href{http://dx.doi.org/10.1103/PhysRevD.68.083519}{{\em Phys. Rev.}
  {\bfseries D68} (2003) 083519},
\href{http://arxiv.org/abs/astro-ph/0309135}{{\ttfamily arXiv:astro-ph/0309135
  [astro-ph]}}.

\bibitem{astro-ph/0403291}
S.~Hannestad, ``{What is the lowest possible reheating temperature?},''
  \href{http://dx.doi.org/10.1103/PhysRevD.70.043506}{{\em Phys. Rev.}
  {\bfseries D70} (2004) 043506},
\href{http://arxiv.org/abs/astro-ph/0403291}{{\ttfamily arXiv:astro-ph/0403291
  [astro-ph]}}.

\bibitem{astro-ph/9903475}
S.~Hannestad, ``{Probing neutrino decays with the cosmic microwave
  background},'' \href{http://dx.doi.org/10.1103/PhysRevD.59.125020}{{\em Phys.
  Rev.} {\bfseries D59} (1999) 125020},
\href{http://arxiv.org/abs/astro-ph/9903475}{{\ttfamily arXiv:astro-ph/9903475
  [astro-ph]}}.

\bibitem{1808.05955}
T.~Brinckmann, D.~C. Hooper, M.~Archidiacono, J.~Lesgourgues, and T.~Sprenger,
  ``{The promising future of a robust cosmological neutrino mass
  measurement},'' \href{http://dx.doi.org/10.1088/1475-7516/2019/01/059}{{\em
  JCAP} {\bfseries 1901} (2019) 059},
\href{http://arxiv.org/abs/1808.05955}{{\ttfamily arXiv:1808.05955
  [astro-ph.CO]}}.

\bibitem{1712.01857}
A.~Boyle and E.~Komatsu, ``{Deconstructing the neutrino mass constraint from
  galaxy redshift surveys},''
  \href{http://dx.doi.org/10.1088/1475-7516/2018/03/035}{{\em JCAP} {\bfseries
  1803} (2018) 035},
\href{http://arxiv.org/abs/1712.01857}{{\ttfamily arXiv:1712.01857
  [astro-ph.CO]}}.

\bibitem{1509.07471}
R.~Allison, P.~Caucal, E.~Calabrese, J.~Dunkley, and T.~Louis, ``{Towards a
  cosmological neutrino mass detection},''
  \href{http://dx.doi.org/10.1103/PhysRevD.92.123535}{{\em Phys. Rev.}
  {\bfseries D92} no.~12, (2015) 123535},
\href{http://arxiv.org/abs/1509.07471}{{\ttfamily arXiv:1509.07471
  [astro-ph.CO]}}.

\bibitem{1803.07561}
S.~Mishra-Sharma, D.~Alonso, and J.~Dunkley, ``{Neutrino masses and beyond-
  Lambda-CDM cosmology with LSST and future CMB experiments},''
  \href{http://dx.doi.org/10.1103/PhysRevD.97.123544}{{\em Phys. Rev.}
  {\bfseries D97} no.~12, (2018) 123544},
\href{http://arxiv.org/abs/1803.07561}{{\ttfamily arXiv:1803.07561
  [astro-ph.CO]}}.

\bibitem{hep-ph/0312154}
S.~Hannestad and G.~Raffelt, ``{Cosmological mass limits on neutrinos, axions,
  and other light particles},''
  \href{http://dx.doi.org/10.1088/1475-7516/2004/04/008}{{\em JCAP} {\bfseries
  0404} (2004) 008},
\href{http://arxiv.org/abs/hep-ph/0312154}{{\ttfamily arXiv:hep-ph/0312154
  [hep-ph]}}.

\bibitem{1507.02623}
J.~Morais, M.~Bouhmadi-López, and S.~Capozziello, ``{Can $f(R)$ gravity
  contribute to (dark) radiation?},''
  \href{http://dx.doi.org/10.1088/1475-7516/2015/09/041,
  10.1088/1475-7516/2015/9/041}{{\em JCAP} {\bfseries 1509} (2015) 041},
\href{http://arxiv.org/abs/1507.02623}{{\ttfamily arXiv:1507.02623 [gr-qc]}}.

\bibitem{1406.2961}
P.~Hernandez, M.~Kekic, and J.~Lopez-Pavon, ``{$N_{\rm eff}$ in low-scale
  seesaw models versus the lightest neutrino mass},''
  \href{http://dx.doi.org/10.1103/PhysRevD.90.065033}{{\em Phys. Rev.}
  {\bfseries D90} no.~6, (2014) 065033},
\href{http://arxiv.org/abs/1406.2961}{{\ttfamily arXiv:1406.2961 [hep-ph]}}.

\bibitem{1310.1774}
K.~S. Jeong, M.~Kawasaki, and F.~Takahashi, ``{Axions as Hot and Cold Dark
  Matter},'' \href{http://dx.doi.org/10.1088/1475-7516/2014/02/046}{{\em JCAP}
  {\bfseries 1402} (2014) 046},
\href{http://arxiv.org/abs/1310.1774}{{\ttfamily arXiv:1310.1774 [hep-ph]}}.

\bibitem{1303.6267}
P.~Di~Bari, S.~F. King, and A.~Merle, ``{Dark Radiation or Warm Dark Matter
  from long lived particle decays in the light of Planck},''
  \href{http://dx.doi.org/10.1016/j.physletb.2013.06.003}{{\em Phys. Lett.}
  {\bfseries B724} (2013) 77--83},
\href{http://arxiv.org/abs/1303.6267}{{\ttfamily arXiv:1303.6267 [hep-ph]}}.

\bibitem{1303.5379}
C.~Brust, D.~E. Kaplan, and M.~T. Walters, ``{New Light Species and the CMB},''
  \href{http://dx.doi.org/10.1007/JHEP12(2013)058}{{\em JHEP} {\bfseries 12}
  (2013) 058},
\href{http://arxiv.org/abs/1303.5379}{{\ttfamily arXiv:1303.5379 [hep-ph]}}.

\bibitem{1302.2516}
T.~Higaki, K.~S. Jeong, and F.~Takahashi, ``{A Parallel World in the Dark},''
  \href{http://dx.doi.org/10.1088/1475-7516/2013/08/031}{{\em JCAP} {\bfseries
  1308} (2013) 031},
\href{http://arxiv.org/abs/1302.2516}{{\ttfamily arXiv:1302.2516 [hep-ph]}}.

\bibitem{1303.0143}
M.~Archidiacono, E.~Giusarma, A.~Melchiorri, and O.~Mena, ``{Neutrino and dark
  radiation properties in light of recent CMB observations},''
  \href{http://dx.doi.org/10.1103/PhysRevD.87.103519}{{\em Phys. Rev.}
  {\bfseries D87} no.~10, (2013) 103519},
\href{http://arxiv.org/abs/1303.0143}{{\ttfamily arXiv:1303.0143
  [astro-ph.CO]}}.

\bibitem{astro-ph/0507544}
S.~Hannestad, A.~Ringwald, H.~Tu, and Y.~Y.~Y. Wong, ``{Is it possible to tell
  the difference between fermionic and bosonic hot dark matter?},''
  \href{http://dx.doi.org/10.1088/1475-7516/2005/09/014}{{\em JCAP} {\bfseries
  0509} (2005) 014},
\href{http://arxiv.org/abs/astro-ph/0507544}{{\ttfamily arXiv:astro-ph/0507544
  [astro-ph]}}.

\bibitem{1212.1472}
M.~C. Gonzalez-Garcia, V.~Niro, and J.~Salvado, ``{Dark Radiation and Decaying
  Matter},'' \href{http://dx.doi.org/10.1007/JHEP04(2013)052}{{\em JHEP}
  {\bfseries 04} (2013) 052},
\href{http://arxiv.org/abs/1212.1472}{{\ttfamily arXiv:1212.1472 [hep-ph]}}.

\bibitem{1111.0605}
J.~L. Menestrina and R.~J. Scherrer, ``{Dark Radiation from Particle Decays
  during Big Bang Nucleosynthesis},''
  \href{http://dx.doi.org/10.1103/PhysRevD.85.047301}{{\em Phys. Rev.}
  {\bfseries D85} (2012) 047301},
\href{http://arxiv.org/abs/1111.0605}{{\ttfamily arXiv:1111.0605
  [astro-ph.CO]}}.

\bibitem{1111.6599}
D.~Hooper, F.~S. Queiroz, and N.~Y. Gnedin, ``{Non-Thermal Dark Matter
  Mimicking An Additional Neutrino Species In The Early Universe},''
  \href{http://dx.doi.org/10.1103/PhysRevD.85.063513}{{\em Phys. Rev.}
  {\bfseries D85} (2012) 063513},
\href{http://arxiv.org/abs/1111.6599}{{\ttfamily arXiv:1111.6599
  [astro-ph.CO]}}.

\bibitem{1309.5383}
{\bfseries Topical Conveners: K.N. Abazajian, J.E. Carlstrom, A.T. Lee}
  Collaboration, K.~N. Abazajian {\em et~al.}, ``{Neutrino Physics from the
  Cosmic Microwave Background and Large Scale Structure},''
  \href{http://dx.doi.org/10.1016/j.astropartphys.2014.05.014}{{\em Astropart.
  Phys.} {\bfseries 63} (2015) 66--80},
\href{http://arxiv.org/abs/1309.5383}{{\ttfamily arXiv:1309.5383
  [astro-ph.CO]}}.

\bibitem{0808.3137}
S.~Pastor, T.~Pinto, and G.~G. Raffelt, ``{Relic density of neutrinos with
  primordial asymmetries},''
  \href{http://dx.doi.org/10.1103/PhysRevLett.102.241302}{{\em Phys. Rev.
  Lett.} {\bfseries 102} (2009) 241302},
\href{http://arxiv.org/abs/0808.3137}{{\ttfamily arXiv:0808.3137 [astro-ph]}}.

\bibitem{0705.2406}
M.~Lattanzi and J.~W.~F. Valle, ``{Decaying warm dark matter and neutrino
  masses},'' \href{http://dx.doi.org/10.1103/PhysRevLett.99.121301}{{\em Phys.
  Rev. Lett.} {\bfseries 99} (2007) 121301},
\href{http://arxiv.org/abs/0705.2406}{{\ttfamily arXiv:0705.2406 [astro-ph]}}.

\bibitem{1511.00975}
E.~Di~Valentino, E.~Giusarma, O.~Mena, A.~Melchiorri, and J.~Silk,
  ``{Cosmological limits on neutrino unknowns versus low redshift priors},''
  \href{http://dx.doi.org/10.1103/PhysRevD.93.083527}{{\em Phys. Rev.}
  {\bfseries D93} no.~8, (2016) 083527},
\href{http://arxiv.org/abs/1511.00975}{{\ttfamily arXiv:1511.00975
  [astro-ph.CO]}}.

\bibitem{Hamann:2013iba}
J.~Hamann and J.~Hasenkamp, ``{A new life for sterile neutrinos: resolving
  inconsistencies using hot dark matter},''
  \href{http://dx.doi.org/10.1088/1475-7516/2013/10/044}{{\em JCAP} {\bfseries
  10} (2013) 044}, \href{http://arxiv.org/abs/1308.3255}{{\ttfamily
  arXiv:1308.3255 [astro-ph.CO]}}.

\bibitem{Hamann:2011ge}
J.~Hamann, S.~Hannestad, G.~G. Raffelt, and Y.~Y. Wong, ``{Sterile neutrinos
  with eV masses in cosmology: How disfavoured exactly?},''
  \href{http://dx.doi.org/10.1088/1475-7516/2011/09/034}{{\em JCAP} {\bfseries
  09} (2011) 034}, \href{http://arxiv.org/abs/1108.4136}{{\ttfamily
  arXiv:1108.4136 [astro-ph.CO]}}.

\bibitem{Hannestad:2005bt}
S.~Hannestad, A.~Ringwald, H.~Tu, and Y.~Y. Wong, ``{Is it possible to tell the
  difference between fermionic and bosonic hot dark matter?},''
  \href{http://dx.doi.org/10.1088/1475-7516/2005/09/014}{{\em JCAP} {\bfseries
  09} (2005) 014}, \href{http://arxiv.org/abs/astro-ph/0507544}{{\ttfamily
  arXiv:astro-ph/0507544}}.

\bibitem{Lesgourgues:2014zoa}
J.~Lesgourgues and S.~Pastor, ``{Neutrino cosmology and Planck},''
  \href{http://dx.doi.org/10.1088/1367-2630/16/6/065002}{{\em New J. Phys.}
  {\bfseries 16} (2014) 065002},
  \href{http://arxiv.org/abs/1404.1740}{{\ttfamily arXiv:1404.1740 [hep-ph]}}.

\bibitem{hep-ph/0202122}
A.~D. Dolgov, ``{Neutrinos in cosmology},''
  \href{http://dx.doi.org/10.1016/S0370-1573(02)00139-4}{{\em Phys. Rept.}
  {\bfseries 370} (2002) 333--535},
\href{http://arxiv.org/abs/hep-ph/0202122}{{\ttfamily arXiv:hep-ph/0202122
  [hep-ph]}}.

\bibitem{1302.1102}
A.~Palazzo, ``{Phenomenology of light sterile neutrinos: a brief review},''
  \href{http://dx.doi.org/10.1142/S0217732313300048}{{\em Mod. Phys. Lett.}
  {\bfseries A28} (2013) 1330004},
\href{http://arxiv.org/abs/1302.1102}{{\ttfamily arXiv:1302.1102 [hep-ph]}}.

\bibitem{1407.0017}
H.~Baer, K.-Y. Choi, J.~E. Kim, and L.~Roszkowski, ``{Dark matter production in
  the early Universe: beyond the thermal WIMP paradigm},''
  \href{http://dx.doi.org/10.1016/j.physrep.2014.10.002}{{\em Phys. Rept.}
  {\bfseries 555} (2015) 1--60},
\href{http://arxiv.org/abs/1407.0017}{{\ttfamily arXiv:1407.0017 [hep-ph]}}.

\bibitem{Abazajian:2017tcc}
K.~N. Abazajian, ``{Sterile neutrinos in cosmology},''
  \href{http://dx.doi.org/10.1016/j.physrep.2017.10.003}{{\em Phys. Rept.}
  {\bfseries 711-712} (2017) 1--28},
  \href{http://arxiv.org/abs/1705.01837}{{\ttfamily arXiv:1705.01837
  [hep-ph]}}.

\bibitem{Kolb}
E.~W. Kolb and M.~S. Turner, {\em {The Early Universe}}, vol.~69.
\newblock 1990.

\bibitem{dr1}
M.~Cicoli, J.~P. Conlon, and F.~Quevedo, ``{Dark radiation in LARGE volume
  models},'' \href{http://dx.doi.org/10.1103/PhysRevD.87.043520}{{\em Phys.
  Rev. D} {\bfseries 87} no.~4, (2013) 043520},
  \href{http://arxiv.org/abs/1208.3562}{{\ttfamily arXiv:1208.3562 [hep-ph]}}.

\bibitem{ghosh}
B.~S. Acharya, M.~Dhuria, D.~Ghosh, A.~Maharana, and F.~Muia, ``{Cosmology in
  the presence of multiple light moduli},''
  \href{http://dx.doi.org/10.1088/1475-7516/2019/11/035}{{\em JCAP} {\bfseries
  11} (2019) 035}, \href{http://arxiv.org/abs/1906.03025}{{\ttfamily
  arXiv:1906.03025 [hep-th]}}.

\bibitem{Bayer:2020tko}
A.~E. Bayer, A.~Banerjee, and Y.~Feng, ``{A fast particle-mesh simulation of
  non-linear cosmological structure formation with massive neutrinos},''
  \href{http://arxiv.org/abs/2007.13394}{{\ttfamily arXiv:2007.13394
  [astro-ph.CO]}}.

\bibitem{DEramo:2018vss}
F.~D'Eramo, R.~Z. Ferreira, A.~Notari, and J.~L. Bernal, ``{Hot Axions and the
  $H_0$ tension},'' \href{http://dx.doi.org/10.1088/1475-7516/2018/11/014}{{\em
  JCAP} {\bfseries 11} (2018) 014},
  \href{http://arxiv.org/abs/1808.07430}{{\ttfamily arXiv:1808.07430
  [hep-ph]}}.

\bibitem{Riess:2016jrr}
A.~G. Riess {\em et~al.}, ``{A 2.4\% Determination of the Local Value of the
  Hubble Constant},'' \href{http://dx.doi.org/10.3847/0004-637X/826/1/56}{{\em
  Astrophys. J.} {\bfseries 826} no.~1, (2016) 56},
  \href{http://arxiv.org/abs/1604.01424}{{\ttfamily arXiv:1604.01424
  [astro-ph.CO]}}.

\bibitem{AguilarArevalo:2010wv}
A.~A. Aguilar-Arevalo~{\it et al.}, ``Event excess in the {MiniBooNE} search
  for $\bar{\nu}_\mu\rightarrow\bar{\nu}_e$ oscillations,''
  \href{http://dx.doi.org/10.1103/physrevlett.105.181801}{{\em Physical Review
  Letters} {\bfseries 105} no.~18, (Oct., 2010) }.
  \url{https://doi.org/10.1103/physrevlett.105.181801}.

\bibitem{Colin:2007bk}
P.~Colin, O.~Valenzuela, and V.~Avila-Reese, ``{On the Structure of Dark Matter
  Halos at the Damping Scale of the Power Spectrum with and without Relict
  Velocities},'' \href{http://dx.doi.org/10.1086/524030}{{\em Astrophys. J.}
  {\bfseries 673} (2008) 203--214},
  \href{http://arxiv.org/abs/0709.4027}{{\ttfamily arXiv:0709.4027
  [astro-ph]}}.

\bibitem{Boyarsky:2008xj}
A.~Boyarsky, J.~Lesgourgues, O.~Ruchayskiy, and M.~Viel, ``{Lyman-alpha
  constraints on warm and on warm-plus-cold dark matter models},''
  \href{http://dx.doi.org/10.1088/1475-7516/2009/05/012}{{\em JCAP} {\bfseries
  05} (2009) 012}, \href{http://arxiv.org/abs/0812.0010}{{\ttfamily
  arXiv:0812.0010 [astro-ph]}}.

\bibitem{Nadler:2020prv}
{\bfseries DES} Collaboration, E.~Nadler {\em et~al.}, ``{Milky Way Satellite
  Census. III. Constraints on Dark Matter Properties from Observations of Milky
  Way Satellite Galaxies},'' \href{http://arxiv.org/abs/2008.00022}{{\ttfamily
  arXiv:2008.00022 [astro-ph.CO]}}.

\bibitem{Das:2020nwc}
S.~Das and E.~O. Nadler, ``{Constraints on the Epoch of Dark Matter Formation
  from Milky Way Satellites},''
  \href{http://arxiv.org/abs/2010.01137}{{\ttfamily arXiv:2010.01137
  [astro-ph.CO]}}.

\bibitem{DePorzio:2020wcz}
N.~DePorzio, W.~L. Xu, J.~B. Muñoz, and C.~Dvorkin, ``{Finding eV-scale Light
  Relics with Cosmological Observables},''
  \href{http://arxiv.org/abs/2006.09380}{{\ttfamily arXiv:2006.09380
  [astro-ph.CO]}}.

\bibitem{abazajian2019cmbs4}
K.~A. {\it et al.}, ``Cmb-s4 decadal survey apc white paper,'' 2019.

\bibitem{Alam_2017}
S.~Alam, M.~Ata, S.~Bailey, F.~Beutler, D.~Bizyaev, J.~A. Blazek, A.~S. Bolton,
  J.~R. Brownstein, A.~Burden, C.-H. Chuang, and et~al., ``The clustering of
  galaxies in the completed sdss-iii baryon oscillation spectroscopic survey:
  cosmological analysis of the dr12 galaxy sample,''
  \href{http://dx.doi.org/10.1093/mnras/stx721}{{\em Monthly Notices of the
  Royal Astronomical Society} {\bfseries 470} no.~3, (Mar, 2017) }.
  \url{http://dx.doi.org/10.1093/mnras/stx721}.

\bibitem{collaboration2016desi}
D.~Collaboration and {\it et al.}, ``The desi experiment part i:
  Science,targeting, and survey design,'' 2016.

\bibitem{Amendola_2018}
L.~Amendola, S.~Appleby, A.~Avgoustidis, D.~Bacon, T.~Baker, M.~Baldi,
  N.~Bartolo, A.~Blanchard, C.~Bonvin, and et~al., ``Cosmology and fundamental
  physics with the euclid satellite,''
  \href{http://dx.doi.org/10.1007/s41114-017-0010-3}{{\em Living Reviews in
  Relativity} {\bfseries 21} no.~1, (Apr, 2018) }.
  \url{http://dx.doi.org/10.1007/s41114-017-0010-3}.

\bibitem{1708.01530}
{\bfseries DES} Collaboration, T.~Abbott {\em et~al.}, ``{Dark Energy Survey
  year 1 results: Cosmological constraints from galaxy clustering and weak
  lensing},'' \href{http://dx.doi.org/10.1103/PhysRevD.98.043526}{{\em Phys.
  Rev. D} {\bfseries 98} no.~4, (2018) 043526},
  \href{http://arxiv.org/abs/1708.01530}{{\ttfamily arXiv:1708.01530
  [astro-ph.CO]}}.

\bibitem{1812.06076}
H.~Hildebrandt {\em et~al.}, ``{KiDS+VIKING-450: Cosmic shear tomography with
  optical and infrared data},''
  \href{http://dx.doi.org/10.1051/0004-6361/201834878}{{\em Astron. Astrophys.}
  {\bfseries 633} (2020) A69},
  \href{http://arxiv.org/abs/1812.06076}{{\ttfamily arXiv:1812.06076
  [astro-ph.CO]}}.

\end{thebibliography}\endgroup
\end{document}